\documentclass{article}
\usepackage{graphicx}
\usepackage{amsmath}
\usepackage{amsbsy}
\usepackage{epstopdf, epsfig}
\usepackage[a4paper, total={6in, 8in}]{geometry}
\usepackage{xspace}
\usepackage{algorithm}
\usepackage{algpseudocode}
\usepackage{booktab}
\usepackage{subfigure}
\usepackage{soul}
\usepackage{hyperref}

\usepackage{siunitx}
\newcommand{\krs}{\ensuremath{k_\textrm{$r\alpha$}}-\ensuremath{S}\xspace}
\newcommand{\co}{CO\ensuremath{_\textrm{2}}\xspace}
\newcommand{\pcs}{\ensuremath{P_\textrm{c}}-\ensuremath{ S}\xspace}

\title{Modeling of relative permeabilities including dynamic wettability transition zones}

\author{Abay M. Kassa$^{1,2}$ \and
  Sarah E. Gasda$^{2}$ \and
  K. Kumar$^{1}$ \and
 \and F. A. Radu$^{1}$}
\date{}
\begin{document}
\maketitle
\noindent ${}^1$ Department of Mathematics, University of Bergen, P. O. Box 7800, 5020 Bergen, Norway.\\[5pt]
${}^2$ NORCE, Nyg{\aa}rdsgaten 112, 5008 Bergen, Norway.\\[5pt]
Corresponding author: Abay M. Kassa (E-mail: abka@norceresearch.no)
\begin{abstract}
Wettability is a pore-scale property that impacts the relative movement and distribution of fluids in a porous medium. There are reservoir fluids that provoke the surface within pores to undergo a wettability change.  This wettability change, in turn, alters the dynamics of relative permeabilities at the Darcy scale. Thus, modeling the impact of wettability change in the relative permeabilities is essential to understand fluids interaction in porous media. In this study, we include time-dependent wettability change into the relative permeability--saturation relation by modifying the existing relative permeability function. To do so, we assume the  wettability change  is represented by the  sorption-based model that is exposure time and chemistry dependent.   This pore-scale model is then coupled with a triangular bundle-of-tubes model to simulate exposure time-dependent relative permeabilities data. The simulated data is used to characterize and quantify the wettability dynamics in the relative permeability--saturation curves. This study further shows the importance of accurate prediction of the relative permeability in a dynamically altering porous medium.  
\end{abstract}

\section{Introduction}
Wettability alteration (WA) plays an important role in many industrial applications such as microfluidics nanoprinting, enhanced oil recovery (EOR), and \co storage \cite{Bonn09, Iglauer2016,  Yu2008, Iglauer2014, RefBlunt}. Wettability refers to the tendency of one fluid over the others to spread on or adhere to a solid surface  \cite{Olugbenga2014, Bonn09} and is defined by the fluid-fluid contact angle (CA). This pore-scale property regulates the distribution of fluids in the pore spaces and controls the relative flow of immiscible fluids in a porous medium  \cite{Anderson1987, BOBEK1958, Olugbenga2014, Bonn09}. This, in turn, impacts constitutive relations in the multi-phase flow systems such as residual saturation, relative permeability, and capillary pressure at the Darcy scale \cite{Ahmed2001, Pentland2011, Iglauer2011, Iglauer2014,  Olugbenga2014}. Investigating and upscaling the impact of WA on the constitutive relations (hereafter constitutive relations refers to the relative permeability- and capillary pressure-saturation relations) is of great importance.   

Wettability is assumed to be static in time and uniform in space. However, wettability is a dynamic process that depends on surface chemistry, composition of fluids,   exposure time,  and reservoir conditions (pressure and temperature) to name a few \cite{Salatheil1973, Treiber1972, Anderson1987, RefHaag17, Jabhunandan95, Buckley1998, Morrow1970}. 
Experiments on crude oil/brine/rock systems have shown that adsorption of active components from the crude oil is able to change the wettability of the sample porous medium from water-wet to intermediate-wet system \cite{Salatheil1973, Anderson1986a, Buckley1998}. It is also hypothesized that the oil reservoir may be more oil-wet than what is observed from the experiment. This is because the adsorption time of the experiment period was much less than the age of the oil in the reservoir. Furthermore, \co is one of the reservoir fluids which contain active components that can provoke the surface within the pores to undergo a WA \cite{Wang13, Bikkina2011, Yang08, Dickson2006, Iglauer2012,  Jung12, Espinoza2010, Farokhpoor2013, Saraji2013,Iglauer2014}.  

Generally, the WA process can have three phases that delineate the transition from initial to final wetting-state conditions, e.g. initial-wet, final-wet, and dynamic-wet. The end (initial and final) wetting conditions are static in time but can be uniform and mixed in space. A mixed-wet condition could be created by rock mineral and exposure history  differences. This is due to the fact that a pore surface exposed to the WA agent may be altered to a new wetting condition, while the unexposed surface keeps the initial wetting state \cite{Kovescek1993, RefBlunt97}. This creates a mixed-wet condition even within a single pore and was observed and explained first by Salathiel et al. \cite{Salatheil1973} in the 1970s. 
Usually, WA is assumed to occur instantaneously and is considered as a function of the WA agent concentration. In some cases, however, the alteration process might take prolonged time in the scale of weeks and months 
\cite{Tokunaga13, RefW,Buckley1998,Powers1996}. In this regard, the dynamic-wet phase can be a function of exposure time in addition to the WA agent concentration.

The WA process may result in a saturation function alteration for subsequent drainage-imbibition displacements and thus cause hysteresis in constitutive relations   \cite{Vives1999, Ahmed2001, Delshad2003, Spiteri2008, Landry2014}. For instance, core-flooding measurements for (supercritical or gas) \co-water system have indicated that WA-induced alteration in the residual saturation and capillary pressure curves despite the fact that they were measured following a standard procedure, i.e., where ``pressure equilibration'' is obtained after each increment in pressure \cite{Plug07, RefWang, Tokunaga13, Tokunaga2013, Wang13, RefKim12}. In these measurements,  a steadily change in capillary pressure function over time was observed. More importantly, the capillary pressure deviation from the initial-wet state curve could not be explained by classical scaling arguments. The instability and gradual change of residual saturation and capillarity through exposure time, in turn, impact the behavior of relative permeabilities. 

The above experiments reveal that standard constitutive models are not well suited to predict relative permeability and capillary pressure under dynamic, long-term WA. 
One alternative is to use mixed-wet model, e.g. Kjosavik et al. \cite{Kjosavik2002} and Lomeland et al.\cite{Lomeland2005}, that capture the static heterogeneity of wettability in the relative permeabilities. The main feature of these models is their flexibility to describe hysteresis and scanning curves caused by a wettability gradient in space. Other alternatives are models designed to handle the instantaneous WA process in the relative permeabilities. The first class of these models involves a heuristic approach that interpolates between the initial and final wetting states in which the WA effect is captured as a coefficient function \cite{Delshad09, Yu2008, Andersen15, Adibhatla05, Sedaghat2019}. Interpolation models are conceptually simple, while the initial and final wetting states are characterized by standard functions, e.g. Brooks-Corey \cite{RefBrooksCorry} or van Genuchten \cite{RefvanGenuchten}). The other approach incorporates the effect of the instantaneous WA into the relative permeabilities  through the residual saturation directly  \cite{RefLashgari}. 
To date,  only Al-Mutairi et al. \cite{RefAl-Mutairi} have considered  the effect of time-dependent WA in both the relative permeabilities and capillary pressure functions explicitly. However, their model does not sufficiently incorporate or upscale the WA processes to core-scale laws.

Appropriate upscaling of the pore-scale time-dependent WA process connected to the capillary pressure function was the subject of our recent work \cite{Kassa2019}. There, WA dynamics were upscaled by introducing a mechanistic time-dependent CA model at the pore-level that was coupled with a cylindrical bundle-of-tubes model and used to simulate capillary pressure curves for drainage and imbibition displacements. The simulated data was used to formulate and quantify a interpolation-based capillary pressure model at the Darcy scale. The new dynamic model resolves the existing interpolation models used in the studies of reservoir simulation \cite{Adibhatla05, Delshad09, Yu2008, Andersen15, Sedaghat2019} by including the dynamics in time and quantifying the pore-scale WA process to the interpolation model in a systematic manner.

One may consider employing a similar approach to \cite{Kassa2019} and an interpolation-type model to capture the pore-scale underpinnings of WA in the relative permeability behaviors. However,  time-dependent WA may impact the capillary pressure and relative permeabilities in different ways. As observed in \cite{Kassa2019},  WA  has a direct impact on the entry pressure in each pore and reflects it at the Darcy scale. Furthermore, a small change in CA exerts a large impact on the dynamics of the capillary pressure function. However, the relative permeability alteration occurs when the WA affects the pore filling/draining orders of pore-sizes. This may lead to a longer exposure time to observe a relative permeability deviation from the initial-wet state curve. Furthermore, unlike the capillary function, the relative permeability curves are constrained between zero and one for any change of wettability. These features of the relative permeability may impact the modeling approach to upscale the pore-scale WA process to the relative permeability behavior.

To our knowledge, a physically reliable model to characterize a prolonged exposure time-dependent WA induced dynamics in the relative permeability behaviors has not been proposed yet. 
This paper revises and extends the approach discussed in \cite{Kassa2019}   to develop a relative permeability model that includes pore-level time-dependent WA processes. Section \ref{ModelApp} summarizes two possible approaches that can be applied to upscale the impact of time-dependent WA in the relative permeabilities. The fluid-fluid CA change is designed as a function of exposure time to the WA agent at the pore level to measure the WA process. This model is coupled with a pore-scale, triangular bundle-of-tubes, model to simulate time-dependent WA induced relative permeability curves. 
These curves are presented in Section \ref{sec:4} and are used to evaluate the modeling approaches hypothesized in Section \ref{ModelApp}. 
\section{Modeling and simulation approach}\label{ModelApp}
A  time-dependent WA may introduce a dynamic term in the relative permeability--saturation (\krs) relationship. This dynamic term can be  measured by   its deviation from the static initial  wetting-state as: 
\begin{eqnarray}
    k_{r\alpha}(\cdot) - k_{r\alpha}^i(S_\alpha) :=f^\text{d}_\alpha(\cdot),
    \label{eq:dynPc}
\end{eqnarray}
or, the dynamic term can be correlated with the parameters of the standard models
\begin{eqnarray}
    k_{r\alpha}(\cdot) =  k_{r\alpha}^i(S_\alpha, a(\cdot), b(\cdot), \ldots), 
    \label{eq:dynPc2}
\end{eqnarray}
where $a(\cdot)$ and $b(\cdot)$ are fitting parameters that change along exposure time, whereas $f_\alpha^\text{d}$ represents the WA induced dynamic component, and the subscript  $\alpha\in \{w,n\}$ represents the wetting and non-wetting phases. 

In this study, we  explore both approaches in Eqs.~\eqref{eq:dynPc} and \eqref{eq:dynPc2} to quantify and characterize $f^{\rm d}_\alpha$ in the relative permeabilities for a system that undergoes a WA.  From Equation \eqref{eq:dynPc}, we propose an interpolation model following our previous work \cite{Kassa2019}, where the dynamic component is designed to interpolate between two end wetting-state curves.    
To obtain an interpolation model, the dynamic component in Eq.~\eqref{eq:dynPc} can be scaled by the difference between  the initial and final wetting-state relative permeability curves. The resulting quantity is non-dimensional and  referred to as the \textit{dynamic coefficient},
\begin{equation}\label{eq:dynPcpc}
 \omega_\alpha\big(k_{r\alpha}^f-k_{r\alpha}^i\big) = f^d_\alpha,
\end{equation}
where the superscript $i$ and $f$ represents relative permeabilities at the initial and final wetting-states respectively. 
This can be substituted into Eq.~\eqref{eq:dynPc} to obtain   dynamic relative permeability models
\begin{equation}
k_{r\alpha} = (1-\omega_\alpha)k_{r\alpha}^{i} + \omega_\alpha k_{r\alpha}^{f},
    \label{eq:dynPc_interp}
\end{equation}
where $\omega_\alpha$, the dynamic coefficient, is responsible for capturing the wettability dynamics at the macroscale.  Similar model to  Eq.~\eqref{eq:dynPc_interp} were employed to include the impact of  instantaneous WA   into the relative permeability curves \cite{Delshad09, Yu2008, Andersen15, Adibhatla05,Sedaghat2019}.

The approach in Eq.~\eqref{eq:dynPc2} relies on a systematic inclusion of the dynamic term $f_\alpha^{\rm d}$ into the relative permeability function through the model parameters. This can be done by formulating  the parameters $a(\cdot)$ and $b(\cdot)$ as a function of exposure time and WA agent in similar fashion as $\omega_\alpha$. 
This approach is motivated by the fact that the parameters in the standard relative permeability models are adjusted to different values when wettability changes from one state to the other. 

Both the initial and final wetting-state curves can be characterized fully by the well-known relative permeability models such as van Genuchten\cite{RefvanGenuchten} or Brooks-Corey \cite{RefBrooksCorry}, Purcell \cite{Kewen2006}, 
the LET model \cite{Lomeland2005} or a model proposed by  Kjosavik et al. \cite{Kjosavik2002}. For the sake of brevity,  we focus on the Brooks-Corey (BC) and LET models in this study. 
The BC relative permeabilities can be derived by integrating the capillary pressure over the capillary tubes \cite{RefXu, Kjosavik2002}. After the integration of the BC capillary pressure, one can obtain relative permeabilities
\begin{equation}\label{eq:bc1}
 k_{rw}^i = S_w^{a_w},~ {\rm and}~k_{rn}^i = (1-S_w^{a_n})(1-S_w)^{m_n},
\end{equation}
for the water-wet system and 
\begin{equation}\label{eq:bc2}
 k_{rw}^{f} = (1-S_w^{a_w})(1-S_n)^{m_w},~ {\rm and}~k_{rn}^{f} = S_n^{a_n},
\end{equation}
for a hydrophobic system (see \cite{Kjosavik2002}), here  $a_w,a_n, m_w,$ and $m_n$ are data fitting parameters in which the subscripts $w$ and $n$ indicate the wetting and non-wetting surfaces. Particularly the $m$'s are known to be tortuosity exponents.  
In 2005, Lomeland et al. \cite{Lomeland2005} have proposed relative permeability models with three parameters L, E, and T  for two-phase flow system. Their correlation models read as:
\begin{equation}
\label{LET}
k_{rw}^i = \frac{ S_{w}^{L_w}}{S_{w}^{L_w} + E_w(1-S_{w})^{T_w}},~{\rm and} ~ k_{rn}^i = \frac{ (1-S_{w})^{L_n}}{(1-S_{w})^{L_n} + E_nS_{w}^{T_n}},
\end{equation} 
where $L_\alpha$, $E_\alpha$, and $T_\alpha$ are data fitting empirical parameters. A detailed description and explanation of the parameters can be found in \cite{Lomeland2005}. 
The important characteristics of the LET model~\eqref{LET} is its flexibility to predict the relative permeability curves for any type of wettability conditions. We note that the parameters, $a$ and $b$, in Eq.~\eqref{eq:dynPc2} are associated with data fitting parameters in Eqs.~\eqref{eq:bc1}-\eqref{eq:bc2} and \eqref{LET}. 
%
%
%
%


The WA induced dynamics in \krs relation should be characterized in  a unified manner in order to evaluate the behaviors of $\omega_\alpha(\cdot)$, $a(\cdot)$, and $b(\cdot)$ along exposure time to the WA agent. 
The WA induced \krs data  can be measured from laboratory experiments. However, this approach is expensive in terms of time. Thus, we follow a theoretical approach in which we simulate  time-dependent \krs data from a pore-scale model. 
\subsection{Pore-scale model description}
We employed a triangular bundle-of-tubes to represent the pore-scale model. 
Note that one can also use a simpler pore-scale model, i.e., a cylindrical bundle-of-tubes model, to characterize the impact of WA on \krs relations. However, polygonal pores are advanced in the way that they can represent physical processes such as the establishment of mixed wettability within a single pore. This may lead to different fluid distributions within a pore, and establishment of non-wetting fluid layers in the corners of the pore space and drainage through layers \cite{Kovescek1993, Hui2000, RefHelland}. 

\begin{figure}[h!]
\center
\includegraphics[scale=0.995]{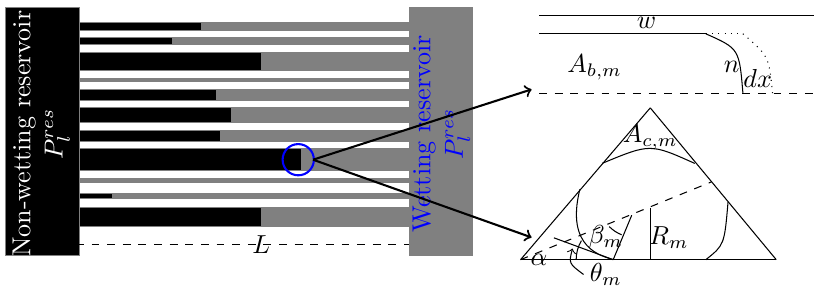}%
\caption{ The fluid displacement scenario in a bundle of tubes that is connected with wetting and non-wetting phase reservoirs. The right column shows fluid distribution during primary drainage. For complete (drainage-imbibition cycles) fluid configurations, see Fig \ref{fig:con}.}\label{Bundfigure}
\end{figure}

A bundle-of-tubes model is  a collection of capillary tubes with a distribution of radii as depicted in Fig \ref{Bundfigure}.
The tubes in Fig  \ref{Bundfigure}  are connected with the wetting (right with pressure $P_{ r}^{res.}$) and non-wetting (left with pressure $P_{  l}^{ res.}$) phase reservoirs. Once the fluid movement is initiated in the tubes, the fluid configurations in each tube can have the form as in Fig~\ref{Bundfigure}. Here,   $A_{b,m}$, $A_{c,m}$, and $dx$ represent the bulk area covered by the non-wetting fluid, the corner area still covered by the wetting phase, and change of fluids along the tube respectively, whereas $\alpha$ is  half angle and $\theta_m$ is fluid-fluid CA. Detailed calculation of these areas and the angle $\beta_m$ is discussed below.

Let the boundary pressures difference be defined as
\begin{equation}\label{rpd}
\Delta P = P^{res}_{l}-P^{res}_{r}, 
\end{equation}
and the tubes in the bundle are filled with the wetting phase initially. To displace the wetting phase fluid in the $m^{\rm th}$ tube, the pressure drop has to exceed the local entry pressure \cite{RefDahle}
\begin{equation}\label{rpd2}
\Delta P > P_{c,m}.
\end{equation}
%
%
If condition (\ref{rpd2}) is satisfied, the non-wetting fluid starts to displace the wetting phase, and the volumetric  flow rate in the  $m^{\rm th}$ can be approximated by the Lucas-Washburn flow model \cite{RefWashburn},  
\begin{equation}\label{eqbundle3}
q_m =\frac{\mathcal{G}_m(R_m,\theta_m)[\Delta P - P_{c,m}]}{8[\mu_{\rm nw} x_m^{int} + \mu_{\rm w}(L-x_m^{int})]}, 
\end{equation}
where, $\mu_{\rm nw}$ and $\mu_{\rm w}$ are non-wetting and wetting fluid viscosities, respectively,   the superscript $int$ stands for fluid-fluid interface,    $q_m = {dx_m^{int}}/{dt}$ is the interface velocity, $R_m$ is the $m^{\rm th}$ tube inscribed radius, and $\theta_m$ is the fluid-fluid contact angle at pore $m$. Here, $\theta_m$ is a general contact angle representation and can be specifically defined as $\theta_{r,m}$ if the interface is receding, $\theta_{a,m} $ if the interface is advancing, and $\theta_{h,m}$ if the interface hinges in the corner of the pores.
The $\mathcal{G}_m$ in Eq.~\eqref{eqbundle3} represents the conductance of the fluids and we follow the work of \cite{Hui2000} to pre-compute the conductance.   The interface is assumed to be trapped when it reaches the outlet of the tube, thus $q_m=0$ when the interface reaches  the boundaries.

\begin{figure}[h!]
\centering
\includegraphics[scale=.45]{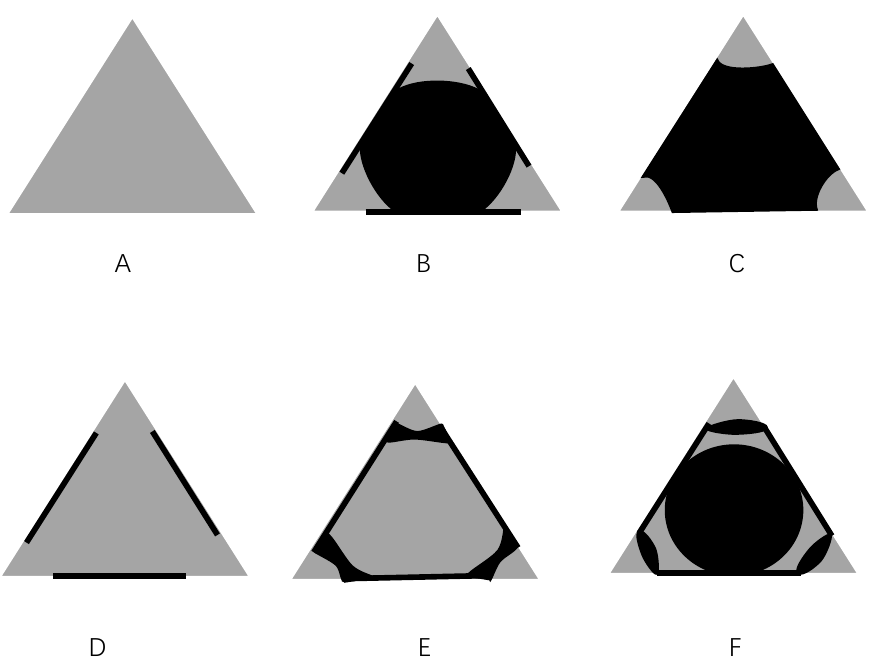}
\caption{Fluid configurations for primary drainage, imbibition, and secondary drainage (water in light color and \co in a dark color). We adopted the configurations from \cite{RefHelland}. The bold lines along the sides indicate altered wettability. } \label{fig:con}
\end{figure}

The wettability change and/or gradient in the polygonal pores may create distinct fluid configurations, for example see Fig~\ref{fig:con}. These configurations can be encountered during drainage (e.g. A, D, and E) and imbibition (e.g. B, C, and F) displacements in each pore in the bundle.  
The bold surface in configurations B through F is to show that part of the surface is exposed to a WA agent at some point and experiences a wettability change. 
Configuration B occurs when the wettability is altered up to a   CA  value satisfying $\theta <\frac{\pi}{2} - \alpha$, where $\alpha = \frac{\pi}{6}$ is the corner half-angle and configuration C otherwise.  
Configuration F may occur when the non-wetting fluid invades configuration E. This is the case when the receding CA is sufficiently smaller than the previous advancing contact angle. 

Below,  relevant aspects of the polygonal pores for the  relative permeabilities measurement  will be stressed, which includes  entry pressure and the areas covered by the two fluids. The    capillary   pressure at the pore-level  is given by
\begin{eqnarray}\label{Gpc}
P_{c,m} = p_n-p_w = \frac{\sigma}{r_m},
\end{eqnarray}
where $p_n$ and $p_w$ are the non-wetting and wetting phase pressures respectively,  $r_m$ is radius of arc meniscus (AMs) that separates the bulk   from the corner fluid, and $\sigma$ is fluid-fluid interfacial tension.
The radius of curvature $r_m$ is determined from the minimization of Helmholtz free energy of the system \cite{Helland2007}. For isothermal, constant total volume, constant chemical potentials, and incompressible system, the minimization of the change in Helmholtz free energy can be simplified to \cite{Helland2007, Morrow1970, Bradford1997} 
\begin{equation}\label{eq:MB}
P_{c,m} dV_n = \sigma(dA_{nw} + \cos(\theta_m) dA_{ns}),
\end{equation}
where $\theta_m$ is fluid-fluid surface angle, $dV_n$ is the change of non-wetting fluid volume, $dA_{nw}$, and $dA_{ns}$ represent the change 
in area of fluid-fluid and fluid-solid interfaces respectively. In \cite{RefHelland, RefDijke06, RefMa96} the entry pressure curvature, $r_m$, and the fluid layer existence criterion were calculated from  Eqs.~\eqref{Gpc} and \eqref{eq:MB} for each of the configurations in Fig \ref{fig:con}. Here, we follow and implement the work of \cite{RefHelland} to calculate the radius of curvature $r_m$ and the associated entry pressure. 

Once we obtained $r_m$ for the $m^{\rm th}$ tube,  fluid volumes (the bulk and corner fluids) can be calculated in terms of $r_m$ and $\theta_m$. To calculate these quantities,   we numbered the fluid-fluid interfaces in order from the apex if there exist more than one interface in the corner, and we apply the indicator notation 
\begin{eqnarray}
I^k =  \left\{\begin{array}{l}
1, ~~ {\rm if~ interface~} k ~{\rm separates~ bulk~ nonwetting~and ~corner ~water}\\
-1,~~ {\rm if ~ interface~} k ~{\rm separates ~bulk~ water~ and ~corner~ nonwetting}.
\end{array}\right.
\end{eqnarray} 
The bulk cross-sectional area   $A_{b,m}^k$   in each tube in the bundle  is defined as,
\begin{equation}\label{bulkarea}
A_{b,m}^k = \left\{\begin{array}{l}
\frac{3R_m^2}{\tan \alpha} - 3r_mb_m^k\sin(\beta_m^k + \alpha) + 3r_m^2\beta_m^k, ~{\rm if} ~I^k = 1,\\[0.1in]
\frac{3R_m^2}{\tan \alpha} - 3r_mb_m^k\sin(\beta_m^k - \alpha)- 3r_m^2\beta_m^k, ~{\rm if} ~I^k = -1,
\end{array}\right.
\end{equation}
where  
\begin{eqnarray}\label{eq:bleng}
b_m^k = \frac{r\sin(\beta_m^k)}{\sin(\alpha)}, ~{\rm and }~ \beta_m^k = \left\{\begin{array}{l} \frac{\pi}{2} -\alpha - \theta_m^k ~  {\rm if} ~I^k = 1\\[0.1in]
 \frac{\pi}{2} + \alpha - \theta_m^k ~  {\rm  if } ~I^k = -1.
\end{array}\right.
\end{eqnarray}
%
Though we used a general notation for CA  $\theta_m^k$  above, $\theta_m^k$ may be replaced by, $\theta_{r,m}$ and $\theta_{a,m}$ when the interface recedes and advances respectively, and $\theta_{h,m}^k$ if the interface, separating the bulk and corner fluids, is hinging. The hinging contact angle (if exists) changes with the entry pressure $P_{c,m}$ according to:
%
\begin{eqnarray}
\theta_{h,m}^k = \left\{\begin{array}{l}
\arccos(\frac{P_{c,m}b_m^k\sin(\alpha)}{\sigma})-\alpha ~  {\rm if} ~I^k = 1\\[0.1in]
\arccos(\frac{P_{c,m}b_m^k\sin(\alpha)}{\sigma})+\alpha ~{\rm  if } ~I^k = -1.
\end{array}\right.
\end{eqnarray}
The fluid area  that occupies  the corner regions can be estimated by 
\begin{equation}\label{cornerarea}
A_{c,m}(\theta_m) = 3r_m^2\Big(\theta_m + \alpha - \frac{\pi}{2} + \cos(\theta_m)\Big(\frac{\cos(\theta_m)}{\tan(\alpha)} -\sin(\theta_m
)\Big)\Big),
\end{equation}
where $\theta_m$ is an argument to determine the appropriate area. The corner surface covered by water in configuration C is obtained by $A_{c,m}(\theta_{h,m}^1)$, whereas the non-wetting fluid layer in configuration E is calculated from $A_{c,m}(\pi-\theta_{a,m}) - A_{c,m}(\theta_{h,m}^1)$. The same approach can be applied if there exist many layers in the corner of the pore.  

As clearly seen above the areas, $A_{b,m}$ and $A_{c,m}$ are   dependent on wettability, i.e., fluid-fluid CA. For example, fluid configuration C and D occurs only if the condition 
$\theta_{m}\leq \frac{\pi}{2} -\alpha$
is satisfied during the drainage displacement. Otherwise, the non-wetting phase occupies the cross sectional area of the tube, and the entry pressure calculation is reduced to the well known Young-Laplace equation. On the other hand, the non-wetting fluid layer occurs in the corner when the condition
$\theta_m > \frac{\pi}{2} + \alpha$
is satisfied. This implies that a dynamic change of CA can also determine the fluid distribution in a single pore. 

\subsection{Pore-scale time-dependent wettability model}\label{sec:3}
Above, we observed that the entry pressure in Eq.~\eqref{eq:MB}, fluid distributions in Eqs.~\eqref{bulkarea}-\eqref{cornerarea}, and a conductance in Eq.~\eqref{eqbundle3} are  wettability evolution dependent. In this section, we introduce a WA mechanism at the pore level to examine its impact on the saturation distribution. We recall that there are many factors that provoke the surface within the pore to undergo a wettability change. Here, we consider the effect of exposure time and fluid-history on the CA change with the assumption that:
\begin{itemize}
\item   The CA of the pore surface is altered through exposure time to the WA agent  and the alteration is permanent. That means the wettability is not restored to the original wetting condition when the WA agent is displaced by the other fluid unless the displacing fluid has a composition that gradually restores the original wetting condition. 

\item  The WA becomes quasi-static in time if the WA agent is removed from the pore before the final wetting-state is reached. If the agent is reintroduced at some later point, alteration continues until the final state.
\end{itemize}
According to the assumptions above, the bulk surface area $A_{b,m}$ is  supposed to be altered dynamically in time, whereas the   corner surface area $A_{c,m}$ may keep the initial condition. Thus, we introduce a functional form to describe a WA mechanism for any arbitrary tube $m$ as
\begin{equation}\label{contactGnera}
\theta_m(\cdot) := \left\{\begin{array}{l}
\theta_{m}^i~~~~~~~~~~~~{\rm for }~A_{c,m},\\[0.06in]
\theta_{m}^i +  \varphi(\cdot){\rm \Delta} {\rm \Theta}~~{\rm for }~A_{b,m}, 
\end{array}\right.
\end{equation}
where ${\rm\Delta  \Theta} = \theta_{m}^f-\theta_{m}^i$, $\theta_{m}^f$, $\theta_{m}^i$ are the final  and initial contact angles  respectively.
The WA model \eqref{contactGnera} is designed to evolve from an arbitrary initial wetting  state to the final wetting condition so that  $\varphi$ is used to interpolate between end wetting conditions  and has a value between zero and one.


Theoretical investigation and detailed laboratory measurement on time-dependent WA is very limited. Furthermore, WA is a complex process, where surface free energies, surface
mineralogy, fluid composition and exposure time interact.  However, adsorption of the WA agent onto the surface area is a natural process  in CA change. 
Such adsorption type wettability evolution is observed for CA measurements \cite{Dickson2006, Jung12, Iglauer2012, Jafri2016,  Davis2003, Morton2005}. These all give an insight to model $\varphi$ in Eq.~\eqref{contactGnera} according to the adsorption of the WA agent and 
can be given as follows
\begin{equation}\label{eq:labda}
\varphi := \frac{\chi_m}{C + \chi_m},
\end{equation}  
where $C$ is a non-dimensional  parameter that controls the speed and extent of alteration from  initial-wet to final-wet system. The derivation of  Eq. \eqref{eq:labda} can be found in our previous work \cite{Kassa2019}. 
The variable $\chi_m$ is a measure of exposure time and is defined as
\begin{equation}\label{eqhistflux2}
\chi_m := \frac{1}{T}\int_0^{t}  \frac{A_{b,m} x_m^{int}}{V_{p,m}} d\tau,
\end{equation}
where $T$ is a pre-specified characteristic time,  $x_m^{int}$ is the fluid-fluid interface position along the tube length $L$, $A_{b,m}$ is the pore surface area that covered by the non-wetting fluid, and $V_{p,m}$ is the pore volume.  Without losses of generality, the characteristic time $T$  is set to be the time for one complete drainage displacement under static initial wetting condition, which can be pre-computed from Eq.~\eqref{eqbundle3}.

For a given interface position $x_m^{int}$, one can  determine the required time to reach the specified interface position from Equation (\ref{eqbundle3}). 
The obtained time is used in Eq.~\eqref{eqhistflux2}  to calculate the exposure history of the  $m^{\rm th}$ pore to the WA agent. According to Eq.~\eqref{eqhistflux2}, each individual pore surface area $A_{b,m}$ will experience CA change based on the   exposure time to the altering fluid, which may be different for different pores depending on the local saturation history. This process would gives rise to a \emph{non-uniform} wetting condition across the bundle  until  all pores have reached the final wetting-state. Moreover, the bulk surface area $A_{b,m}$ is the only surface that undergoes WA, and the corner surface area that covered by water is not subject to WA. This results in  mixed-wet condition. However, this paper do not consider the wettability gradient across the length of the tubes because the time to drain is assumed to be fast compared to the exposure time for WA to occur.



%
%
\subsection{Simulation approach} 
\label{sub:solution_method}
The  wettability dynamics described in Eq.~\eqref{contactGnera} are coupled into  a triangular bundle-of-tubes model to simulate relative permeability curves according to  Algorithm~\ref{algorithm1}. Here, the relative permeability for phase $\alpha$ is calculated as
\begin{equation}
k_{r\alpha} = \frac{Q_\alpha\mu_\alpha L}{\mathcal{K}A_T\Delta P_\alpha},
\end{equation} 
where $Q_\alpha$ is the volumetric total flow rate of phase $\alpha$, $\mathcal{K}$ is absolute permeability  of the bundle, $A_T$ is the cross-sectional area of the bundle.  
%
\begin{algorithm} 
    \centering
    \caption{A single drainage-imbibition cycle. Fluid and rock properties are given according to Table \ref{tab:paralist}}	\label{algorithm1}
    \begin{algorithmic}[1]    \\
        \State \text{\textbf{Drainage displacement}}\\
        \State \text{Set $P_c^{\rm max}$ \% the maximum capillary pressure}
        \While{$\Delta P < P_c^{\rm max}$}         
        \State Calculate $P_c$ from Eq.~\eqref{eq:MB} 
        \If{$\Delta P > P_c$}
        \State \text{Calculate ~ $S_{nw}$ ~ and ~} 
        %
        $\overline{\chi} = \frac{1}{T}\int_0^tS_{nw} d\tau$
        %
        \State \text{Calculate $K_{r\alpha}$}
        \State \text{Update $\theta_m$ from Eqs. \eqref{contactGnera} and \eqref{eqhistflux2}}   
        \EndIf     
        \State Increase $\Delta P$
        \EndWhile\\ 
        %
        %
        \State  \text{\textbf{Imbibition displacement}}\\
        \State  \text{Set minimum entry pressure $P_c^{\rm min}$}
        \While{$\Delta P > P_c^{\rm min}$}         
        \State Calculate $P_c$ from Eq.~\eqref{eq:MB} 
        \If{$\Delta P < P_c$}
        \State \text{Calculate ~ $S_{nw}$ ~ and $\overline{\chi}$} 
        \State \text{Calculate $K_{r\alpha}$}
        %
           
        %
        \State \text{Update $\theta_m$ from Eqs. \eqref{contactGnera} and \eqref{eqhistflux2}}   
        \EndIf     
        \State Increase $\Delta P$
        \EndWhile
    \end{algorithmic}
\end{algorithm}
%
%
We have repeated algorithm \ref{algorithm1} for a few numbers of drainage-imbibition cycles provided that the WA process is completed within these displacements. The flow rate (or the pressure drop $\Delta P$) is controlled in  an arbitrary manner in order to gain $\theta^{\rm f}$ for each tube within a few numbers of drainage-imbibition cycles. 
When we reduce the increment of each $\Delta P$, the flow rate becomes very slow. This imposes a prolonged exposure to the WA agent and thus $\chi_m$ grows and eventually results in a large change in CA. For example, each $\Delta P$  increment is reduced by  three order of magnitude for the last cycle compared to the first cycles to complete the alteration process. 


The relative permeabilities--saturation ``data points'' are obtained in each drainage-imbibition cycle, and the obtained data is presented in the following Section. 
The generated \krs curves   are used to quantify the dynamics  in the relative permeabilities which caused by time-dependent WA. The goal is to develop a correlation model that involves only a few parameters. Finally, the relation between these  parameters and changes in the pore-scale WA model parameter $C$ is studied and examined.

\section{Simulation results}\label{sec:4}
The two-phase flow simulation tool at the pore-scale (Algorithm~\ref{algorithm1}) is implemented in MATLAB. The pore-scale model consists of parallel triangular tubes that connect the non-wetting and wetting reservoirs. Each tube in the bundle is assigned a different radius R, with the radii drawn from a truncated two-parameter Weibull distribution \cite{Hui2000} 
\begin{equation}
 R  = (R_{\rm max} - R_{\rm min})\Big\{-\delta\ln\Big[x(1-{\rm exp}(-1/\delta)) + {\rm exp}(-1/\delta)\Big]\Big\}^{1/\gamma} + R_{\rm min},\label{eq:weibull}
\end{equation}
where $R_{\rm max}$ and $R_{\rm min}$  are the pore radii of the largest and  smallest  pore sizes respectively,   and $\delta$ and $\gamma$ are dimensionless parameters.  The rock parameters and fluid properties are listed in Table \ref{BtubeM}.
\begin{table}[!ht]
\centering
\begin{tabular}{l l l l l l}
\toprule
parameters             & values & unit & parameters  & values & unit\\
\midrule
$\sigma$          & 0.0072 & {\rm N/m}      & no. radii               & 500    & [-]\\
$R_{\rm min}$ & 1    & {\rm  $\mu$m}       & $R_{\rm max}$          &  100   & {\rm  $\mu$m}\\
$\theta_m^f$    & 180    &    degree        & $\theta_m^i$ & 0.0    &    degree\\
$\mu_{\rm w}$ & 0.0015 & {\rm Pa.s}     & $\mu_{\rm nw}$         & 0.0015 & {\rm Pa.s}\\
   L             & 0.001    & {\rm  m} \\
$\delta$  & 1.5 & [-]&  $\gamma$ & 0.5 & [-]\\
\bottomrule
\end{tabular}
\caption{ Parameters used to simulate quasi-static fluid displacement in a bundle-of-tubes. }\label{BtubeM}
\end{table}
These parameters are coupled to the bundle-of-tubes model to simulate fluid conductance and relative permeability curves. 
In the following section, we present and discuss the simulated relative permeability and related results. 

\subsection{End wetting-state relative permeability}\label{staticCorelation}
\begin{figure}[t!]
\centering
\subfigure[]{\includegraphics[scale=0.4]{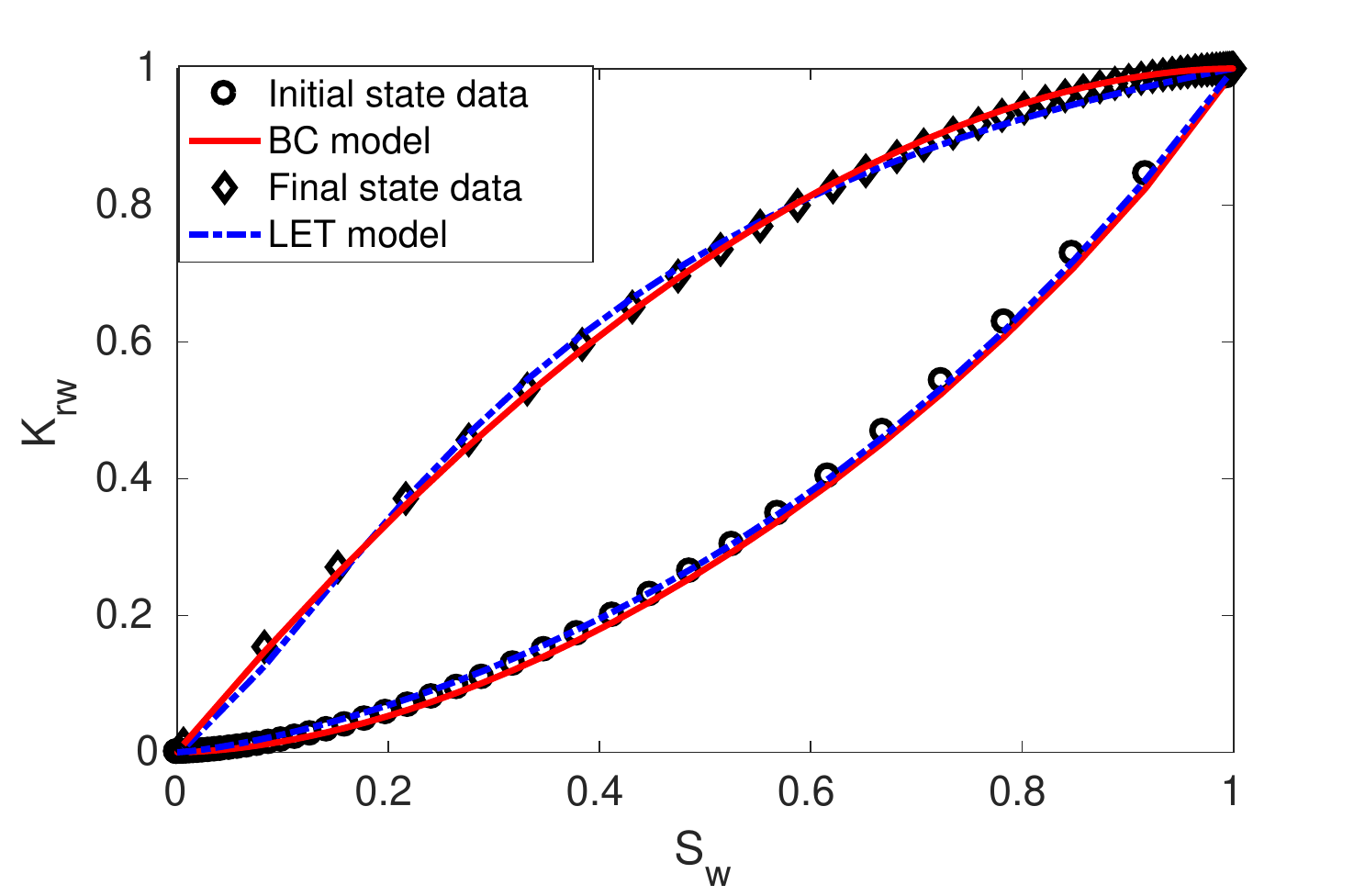}}
\subfigure[]{\includegraphics[scale=0.4]{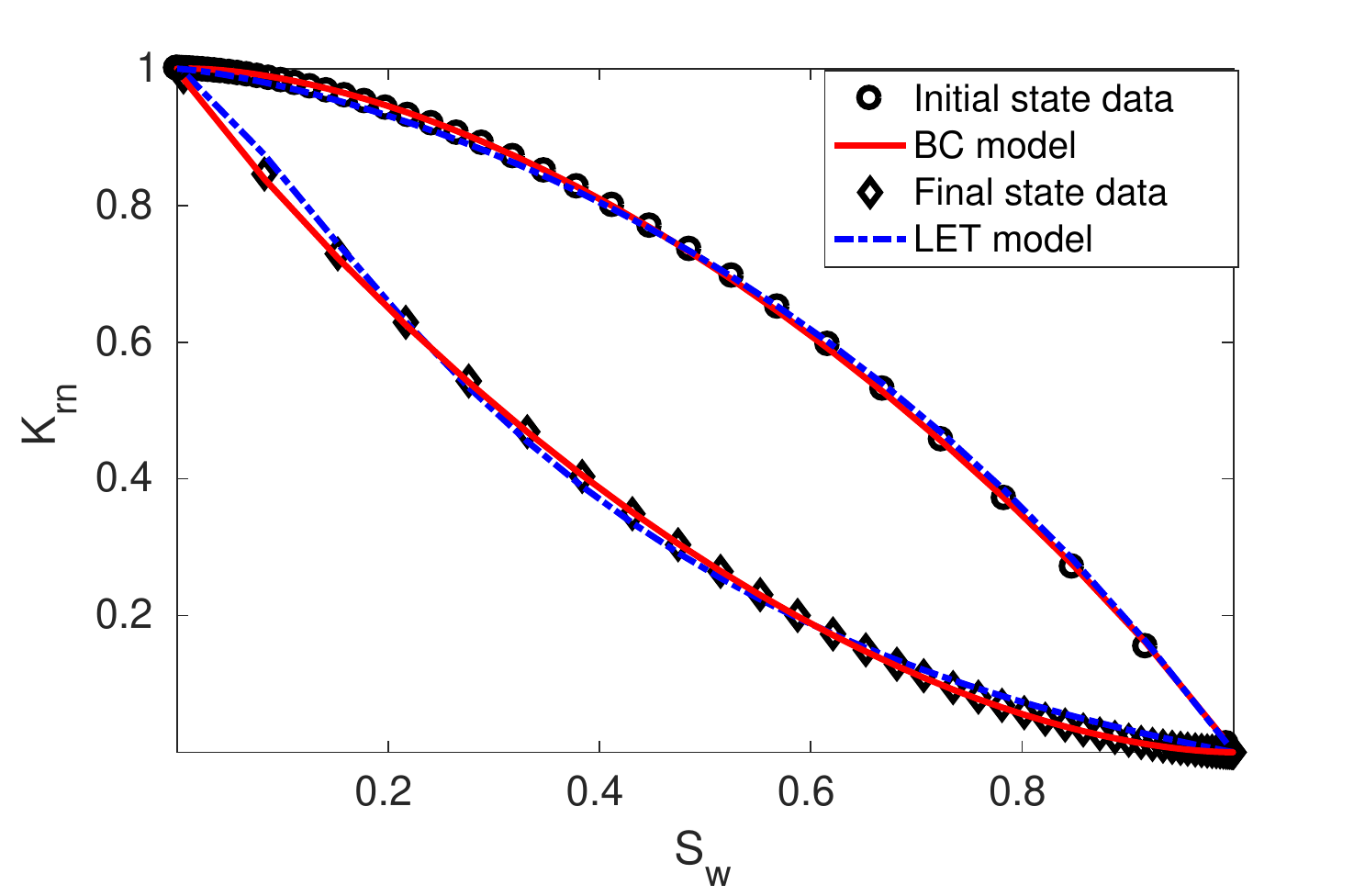}}
\caption{The BC and LET correlation models compared to the initial and final wetting-state relative permeability curves: (a) wetting phase and (b) non-wetting phase relative permeabilities. 
}\label{fig:corre1}
\end{figure}

%
In section \ref{ModelApp}, we point-out that end wettingstate relative permeabilities are the foundation to characterize the dynamic relative permeability curves. Thus, it is natural to examine the end-state \krs relations before quantifying the dynamic relative permeabilities using the same pore-size distribution and fluid properties   in Table \ref{BtubeM}. Thus, we simulated static \krs data by fixing the wettability at pre-specified initial $\theta_m^{i}$ and final $\theta_m^f$ values   in each tube, see Table~\ref{BtubeM}. The simulated curves are plotted along the saturation path in Fig~\ref{fig:corre1}.

We correlated both the BC (in Eqs.~\eqref{eq:bc1}-\eqref{eq:bc2}) and   LET  (in  Eq.~\eqref{LET})  models with the simulated  (initial and final wetting-state) static \krs curves, and the result is compared in Fig~\ref{fig:corre1}.  
\begin{table}[h!]
\centering
\begin{tabular}{l l  l l l  l}
\toprule
Model         & Parameters  & Initial value ($\theta^i$)  &    Final value ($\theta^f$)    \\
\midrule
              & $a_w$    &  0.9         &  1.839     \\ 
BC              & $m_w$   &  0.7329      &  0   \\ [0.05in]
              & $a_n$    &  1.65       &  0.908    \\ 
              & $m_n$   &  0.075       &  0.7329   \\ 
              \midrule
           & $L_w$     &  1.3           &  1.3     \\ 
           & $E_w$     &  2.08          & 0.3   \\
 LET      & $T_w$     &  1            & 1\\ [0.05in]
           
           & $L_n$ & 1          &   1\\
     & $E_n$  &  0.48             & 3.37    \\ 
           & $T_n$  &  1.3        & 1.3   \\ 
 \bottomrule
\end{tabular}
\caption{ Estimated correlation parameter values for initial and final wetting-state relative permeability curves. 
}\label{tab:paralist}
\end{table}
The fitted parameters for the correlation models are found in  Table \ref{tab:paralist}.  
Both the BC correlations and LET models give an excellent match to the simulated phase relative permeability curves under static conditions.

As a complement, we have done experiments on the sensitivity of model parameters (not shown here) for different pore-size distributions by setting  the wetting condition to be static. In this experiment, we found that the parameters $L_n$ and $T_w$ in the LET model can be unity for any pore-size distribution. Furthermore,  from the end wetting-state correlation results, we observe that $E_\alpha$ is the only parameter that depends on wettability change, whereas the BC model parameters are sensitive for wettability change, see Table \ref{tab:paralist}.  More importantly, we have noticed that the parameters $L_w$ and $T_n$ have the same value and thus, we can represent them with a single parameter (say $\lambda$).  Furthermore, the parameter $E_n$ is the inverse of $E_w$. In this regard, we can reduce the LET model  to a two-parameter (but in each phase) model and we call it \textit{reduced LET} model which read as
\begin{equation}
\label{LET_reduced1}
k_{rw} =  E_nS_w^{\lambda}\big(E_nS_w^{\lambda} + 1-S_w\big)^{-1},~ {\rm and }~ k_{rn} = (1-S_w)\big(1-S_w + E_nS_w^{\lambda}\big)^{-1}.
\end{equation}
The model in Eq.~\eqref{LET_reduced1} is a two parameter model with only one parameter that varies along the wettability change. This makes the reduced LET model more reliable than its counter part, i.e., the BC model. Note that the parameter $\lambda$ can also vary with wettability. In this case, a better match with the simulated relative permeability data can be obtained. 

%
%
%
\subsection{Simulated relative permeabilities} 

\begin{figure}[h!]
\centering
\subfigure[]{\includegraphics[scale=0.45]{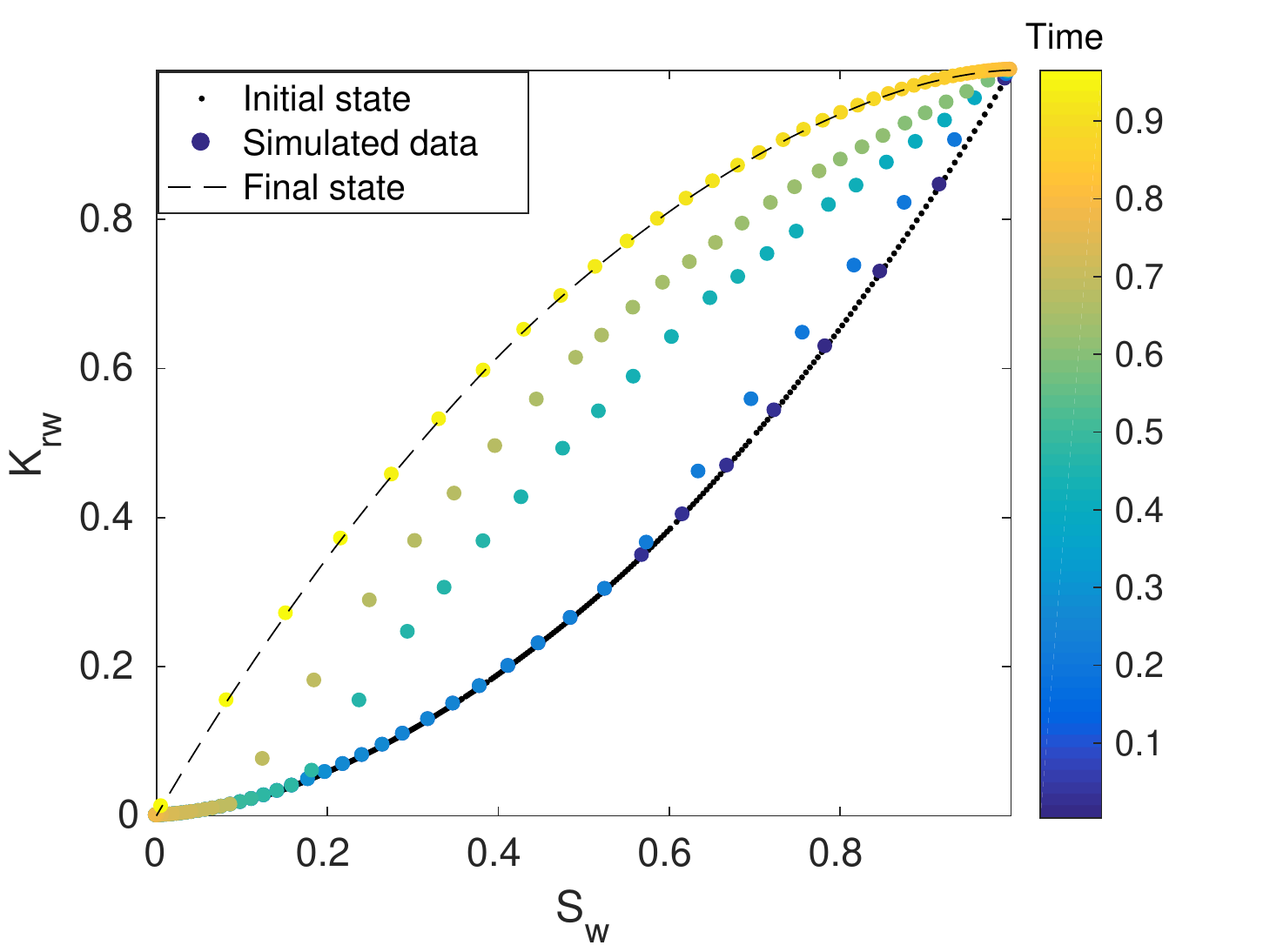}}
\subfigure[]{\includegraphics[scale=0.45]{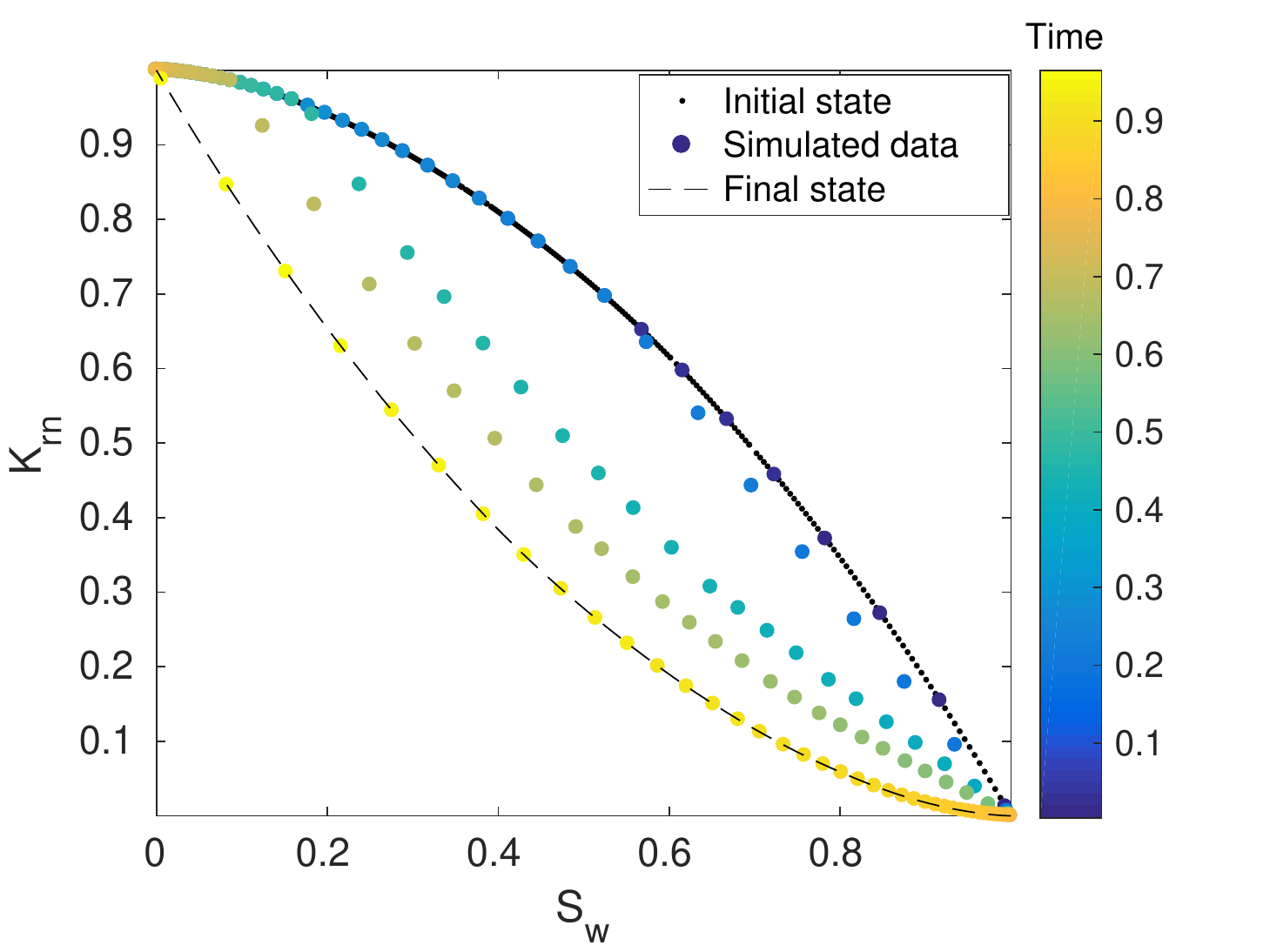}}
\caption{Simulated dynamic relative permeability curves ((a) wetting phase and (b) non-wetting phase) with respect to wetting phase saturation.  The static \krs curves for the initial and final wetting states are shown as a reference. The color code shows the \krs dynamics within a year of exposure time. 
}\label{fig4:4s}
\end{figure}

\begin{figure}[t!]
\centering
\includegraphics[scale=.65]{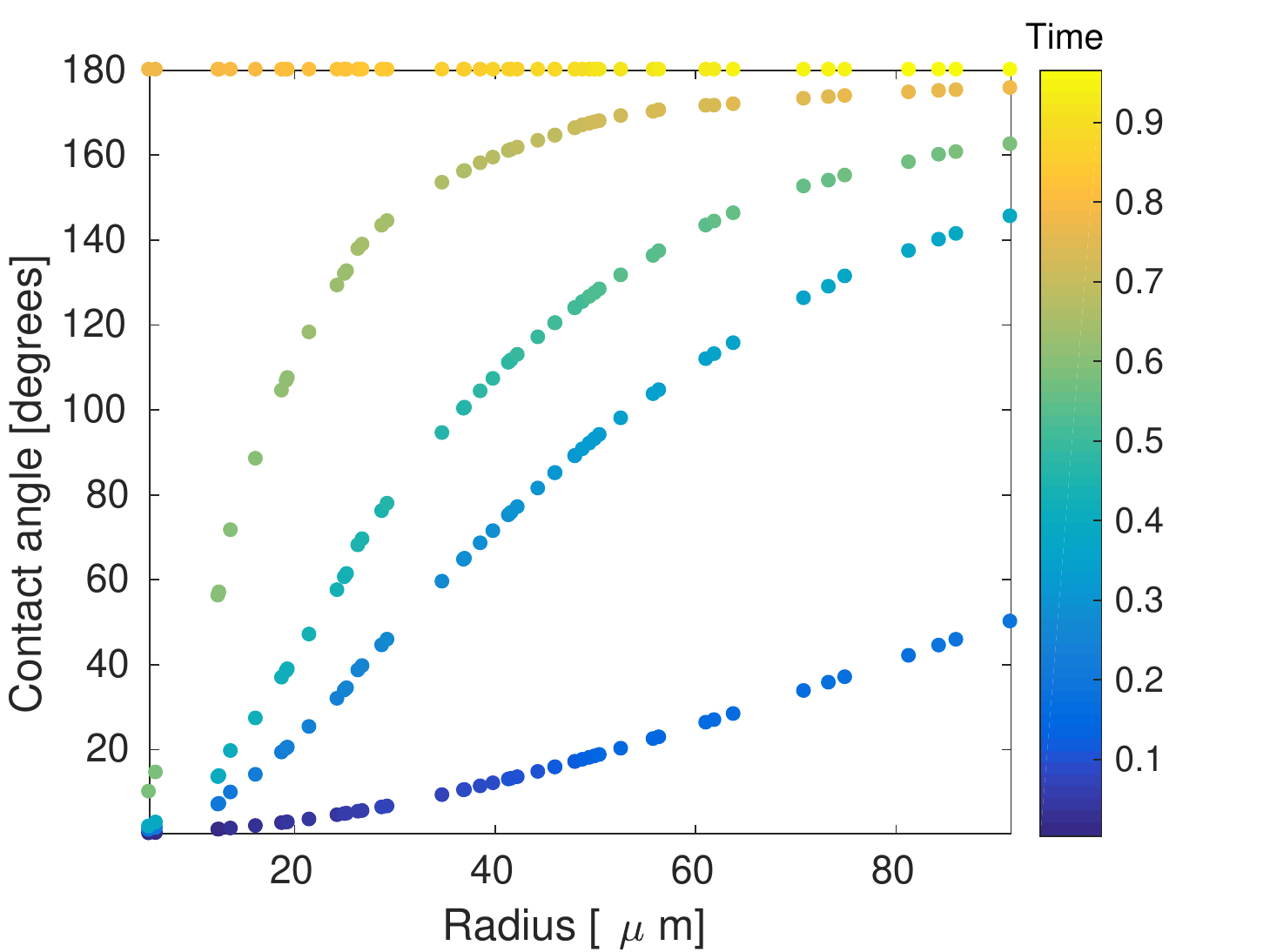}
\caption{Dynamic CA evolution as a function of exposure time to the WA agent per each tube. This CA distribution was recorded at the end of each drainage-imbibition cycle and the color code shows the CA dynamics during  a year of exposure time.}\label{fig4:4} 
\end{figure}

We demonstrate five drainage-imbibition cycles and simulate dynamic \krs relations in each cycle (see Fig~\ref{fig4:4s}), while the tubes in the bundle are altered through time (see Fig~\ref{fig4:4}) following the CA model \eqref{contactGnera}.  These data are generated with a pore-scale parameter {$C = 10\times 10^{-5}$}. Note that the CA change may be halted (temporarily) in the pores if the displacement is to configuration D after imbibition. In this case, the drainage curve may follow the previous imbibition path. However, the imbibition curve may show a deviation from the previous drainage curve if wettability is altered sufficiently. %
%

According to the  CA distribution in Fig~\ref{fig4:4} and \krs curves in Fig~\ref{fig4:4s}, the wettability that ranges from strongly to weakly water-wet does not affect the pore filling and draining orders of the pore-sizes. As a consequence, the \krs relations for drainage/imbibition displacements follow the same saturation path of the initial-wet condition. This implies that a WA induced \krs hysteresis may not occur in a straight bundle-of-tubes model for a wettability range of $\ang{0}\leq \theta_m\leq \ang{60}$. Similar \krs relations are reported for a pore-network model with a wettability range of $\theta_m = \ang{0}$ to $\ang{45}$ \cite{Ahmed2001}. This is in contrast to the capillary pressure--saturation relation, where a small change of CA impacts the \pcs path significantly \cite{Kassa2019}. However, imbibition/drainage displacement may not necessarily occur in monotonically increasing/decreasing order of pore-sizes when the wettability of the pores (some) are altered to intermediate/weakly hydrophobic. This results in a relative permeability--saturation path deviation as observed in Fig~\ref{fig4:4s}. 

We observe that   the wetting and non-wetting phase relative permeabilities steadily increase and decrease  (see Fig \ref{fig4:4s}) respectively in each subsequent drainage-imbibition displacements when the wettability evolves from  hydrophilic to hydrophobic conditions. This occurs because the relatively larger pores start allowing the water (originally wetting phase) to imbibe through before the smaller pores when the medium is altered to intermediate/hydrophobic system. On the contrary, the originally non-wetting fluid prefers the smaller pores to flow in during drainage. As a consequence, a mobility reduction and improvement respectively for the non-wetting and wetting phase relative permeabilities are observed.  However,  any additional drainage-imbibition cycle would follow along the static curve for the final wetting state once the final CA ($\theta^f$)   is reached. 
\begin{figure}[t]
\centering
\subfigure[]{\includegraphics[width=0.48\textwidth]{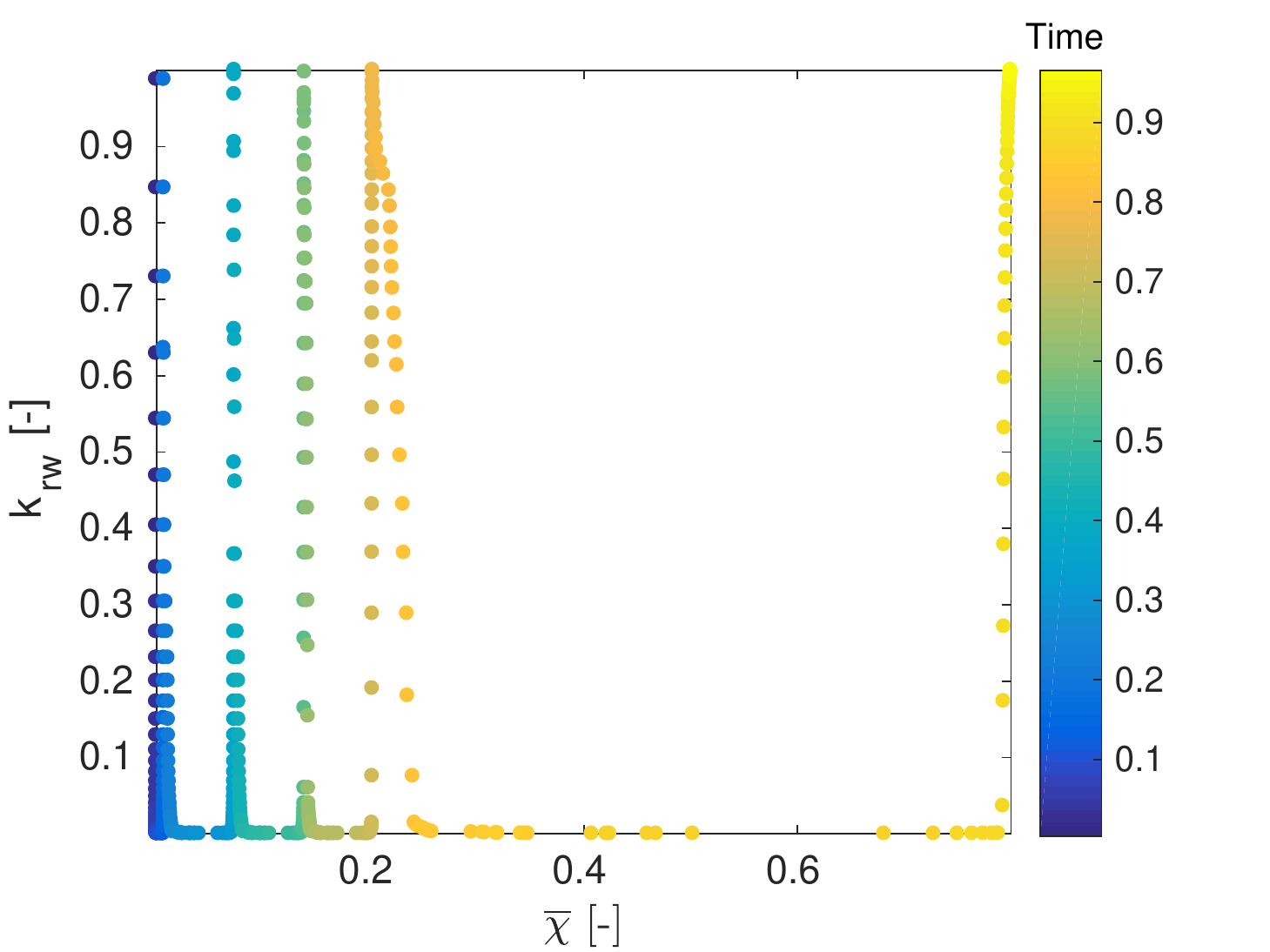}}
\subfigure[]{\includegraphics[width=0.48\textwidth]{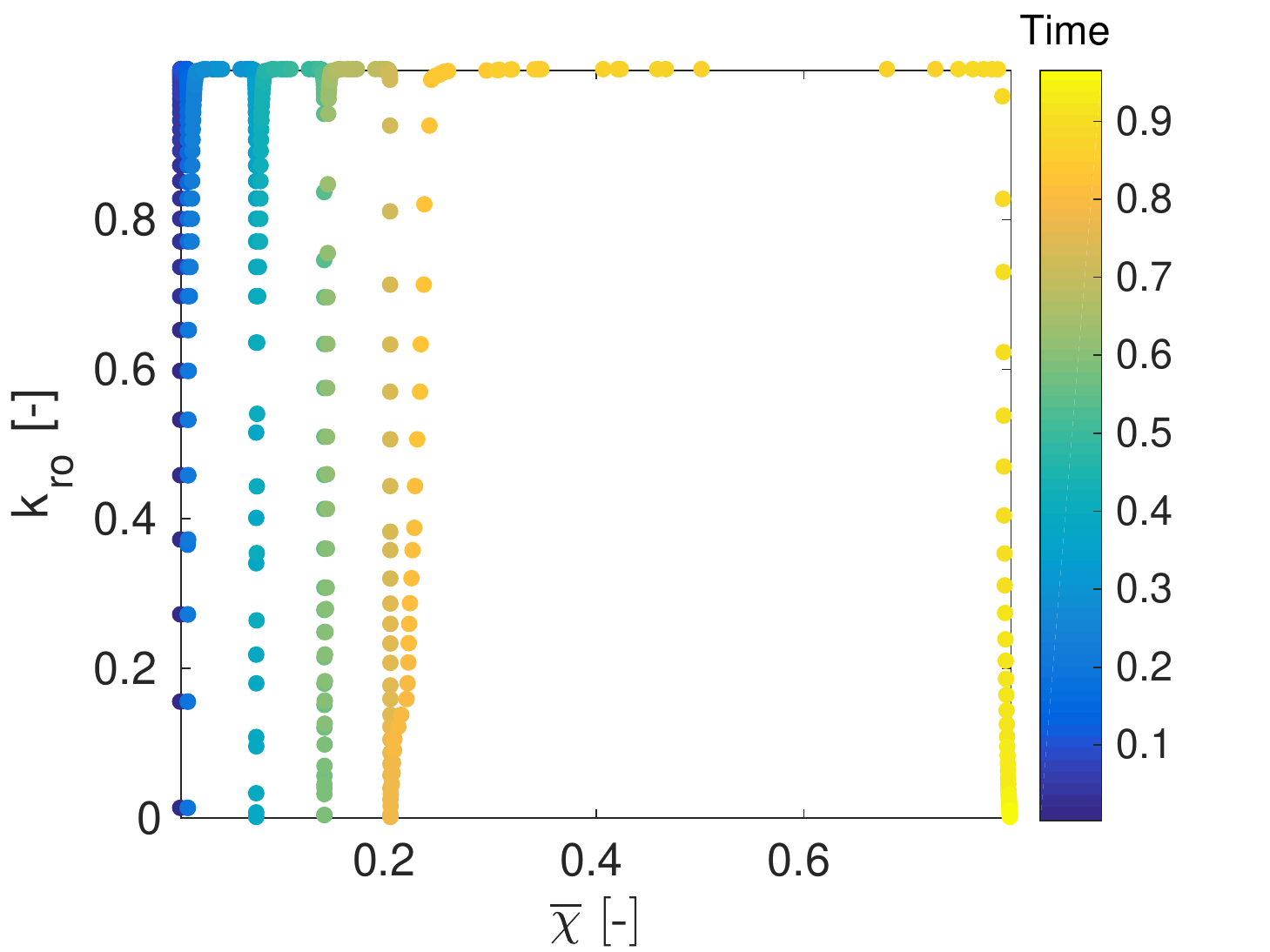}}
\caption{The relative permeability data: (a) wetting phase relative permeability and   (b) non-wetting phase relative permeability as a function of  $\overline{\chi}$ for the non-uniform WA case. The color of each data point indicates the time elapsed in years.}\label{fig4:1b} 
\end{figure}

Analogous to the capillary pressure--saturation relation \cite{Kassa2019}, time-dependent WA introduces dynamic hysteresis in the \krs relations for a bundle-of-tubes model  as shown in Fig~\ref{fig4:4s}. 
The \krs hysteresis imposes a non-unique relation between the relative permeabilities and saturation. This is one of the challenging features of  WA during the quantification of the dynamics in relative permeabilities. To eliminate the hysteresis observed in the \krs relation, we projected the simulated relative permeability data onto the temporal domain $\overline{\chi}$ in Fig~\ref{fig4:1b}. 
The temporal domain, $\overline{\chi}$ is the measure of the exposure history in averaged sense which defined as:
\begin{equation}
\overline{\chi} = \frac{1}{T}\int_0^tS_{nw}d\tau.
\end{equation} 
%
Unlike \krs, $k_{r\alpha}$-$\overline{\chi}$ is uniquely related but non-monotonically i.e., it raises to one and decreases to zero in time along each drainage-imbibition cycle.   
%
%

The WA process also affects the corner fluid distribution in each drainage/imbibition displacement. 
%
\begin{figure}[t!]
\centering
\subfigure[]{\includegraphics[width=0.48\textwidth]{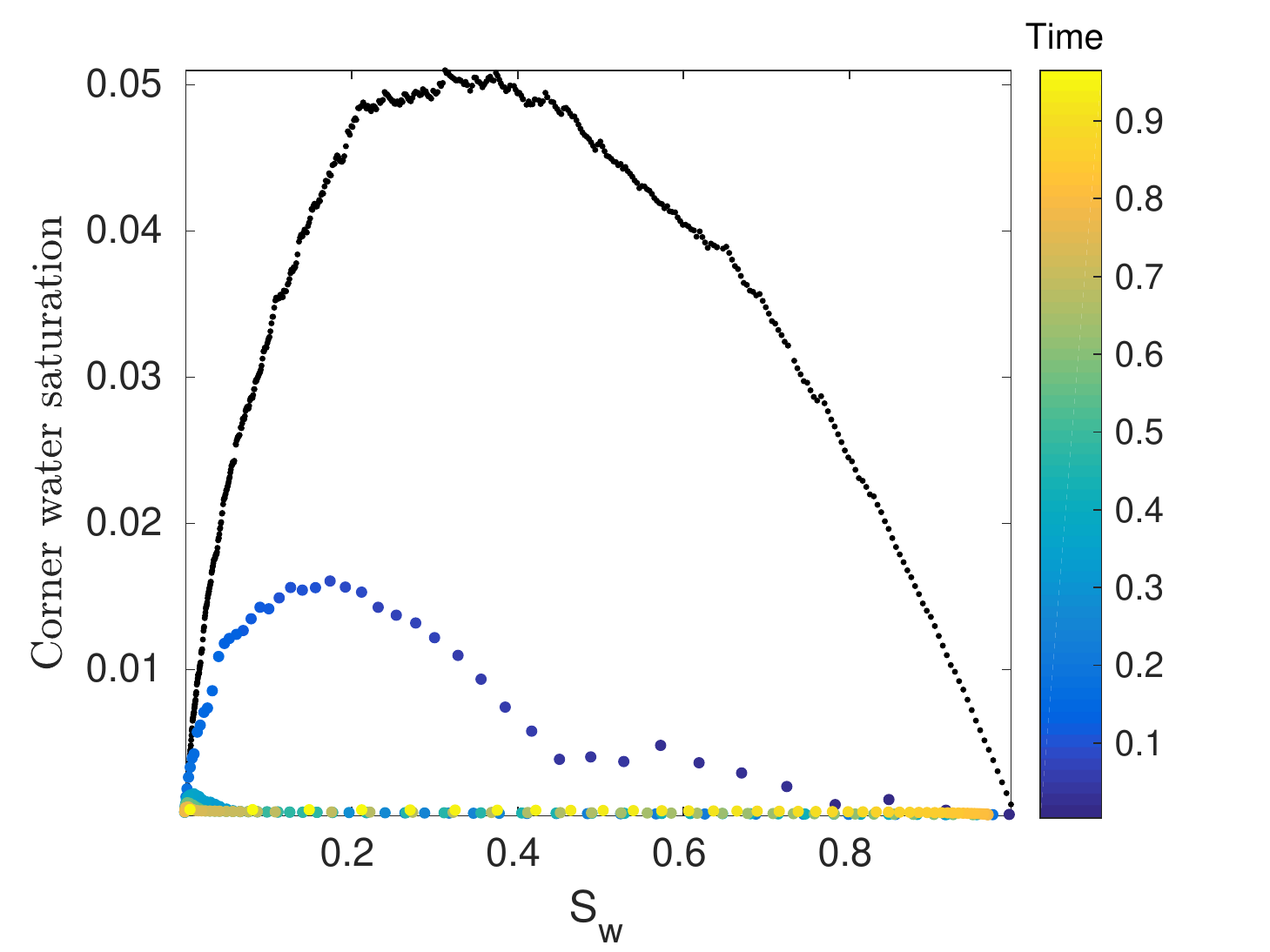}}
\subfigure[]{\includegraphics[width=0.48\textwidth]{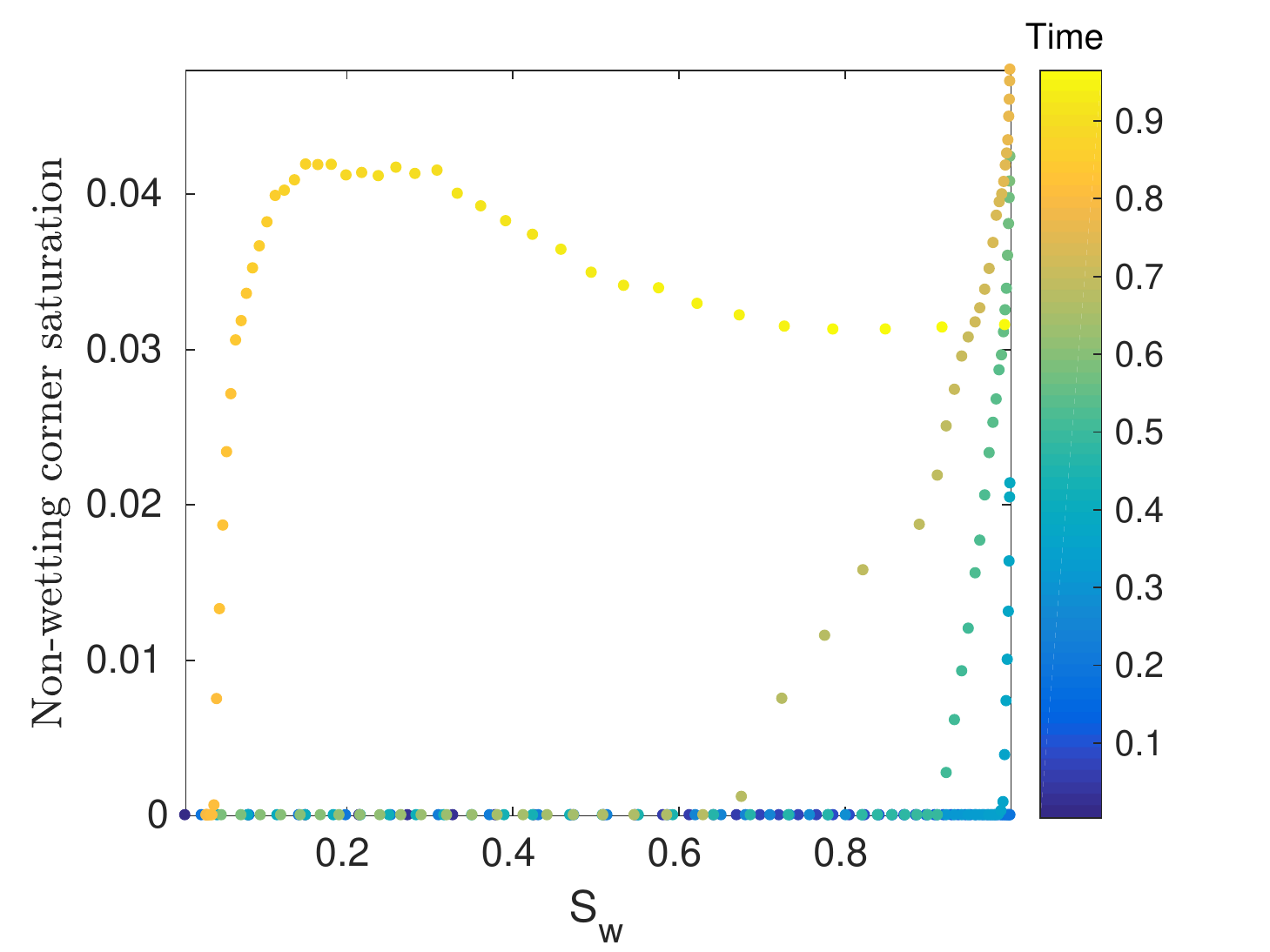}}
\caption{Corner fluid saturations: (a) wetting phases saturation and   (b) non-wetting phase saturation The color of each data point indicates the time elapsed in years.}\label{fig4:1b} 
\end{figure}
%
Fig~\ref{fig4:1b} shows the evolution of (average)  corner wetting and non-wetting  saturations. The wetting phase saturation decreases through time while the layer saturation grows. This is because wettability is altered from water-wet to hydrophobic and thus, the originally non-wetting fluid prefers   to 
be in the corners. However, when (all) pores become more hydrophobic, the corner water increases and bulges-out during imbibition. This process may result in a layer collapse and has shown in the fifth imbibition, where the layer saturation increases during larger pores were imbibed and decrease when smaller pores imbibed over exposure time.  Here, we have checked the establishment of corner fluids   during the calculation of \krs relations in each drainage-imbibition displacements. 

%
%
 
%

%
\subsection{Dynamic relative permeability model development} \label{modeling}
%
%
%
%
%
%

\begin{figure}[h!]
\subfigure[]{\includegraphics[scale=.45]{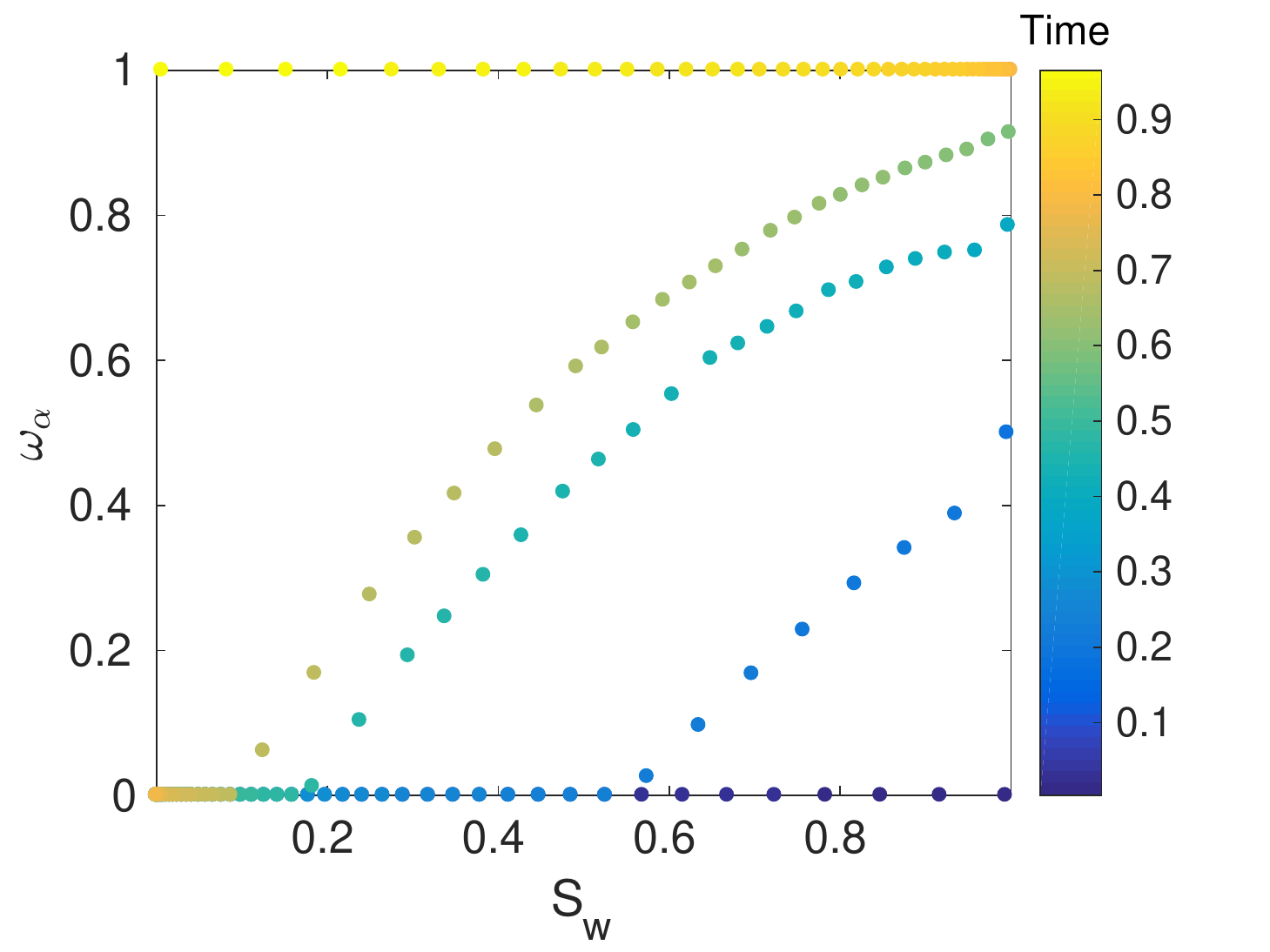}}
\subfigure[]{\includegraphics[scale=.45]{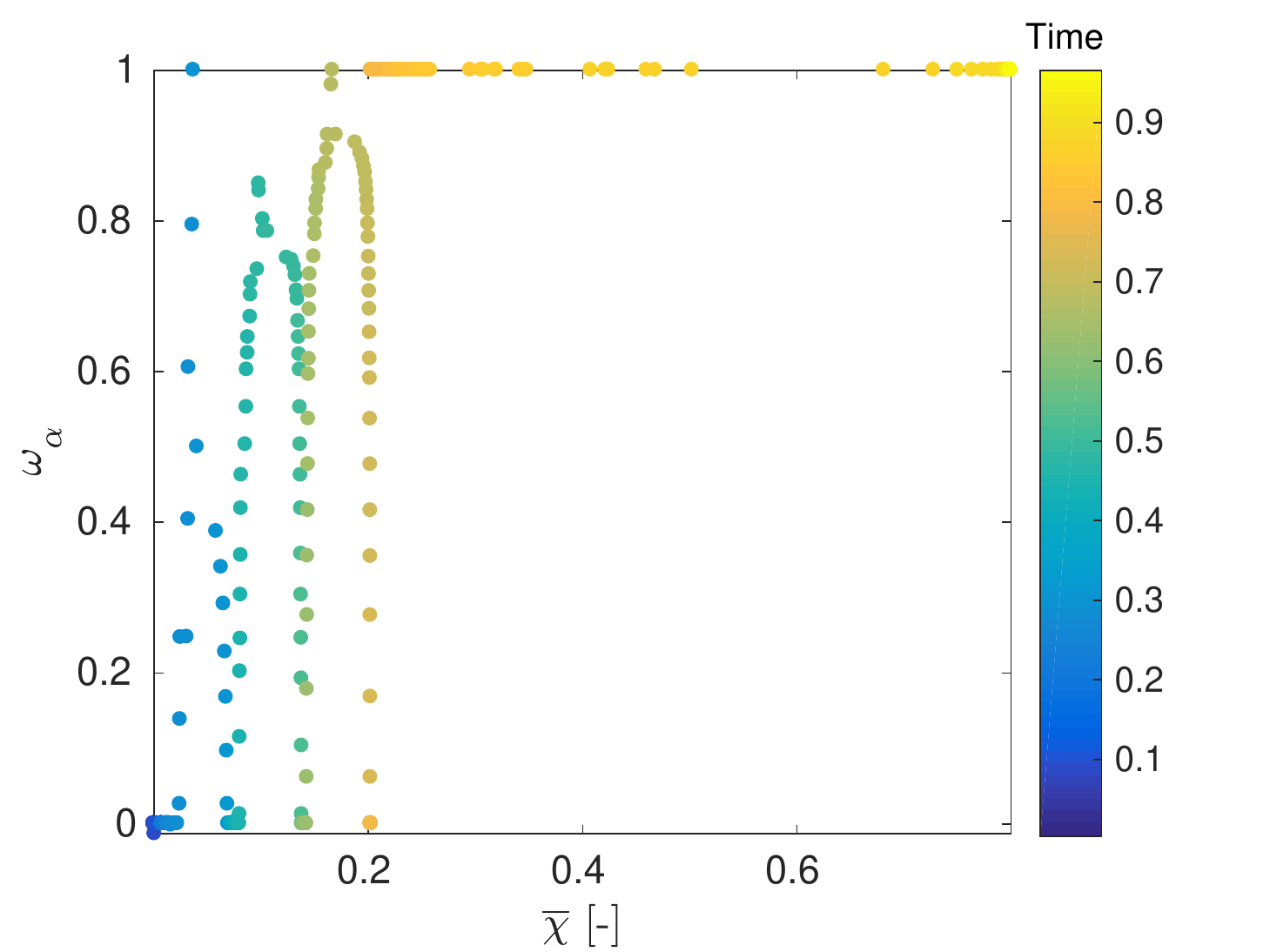}}
\caption{The scaled dynamic deviation from the initial wetting-state relative permeability curves as a function of wetting phase saturation (a) and exposure time $\overline{\chi}$ (b).}\label{fig4:4dif2}
\end{figure}

The interpolation approach, similar to Eq.~\eqref{eq:dynPc_interp}, was applied successfully to capture time-dependent WA mechanisms in the capillary pressure curves \cite{Kassa2019}. Thus, it is important to test the potential of the interpolation-based model (in Eq.~\eqref{eq:dynPc_interp}) to predict time-dependent dynamics in \krs relation. 
The  dynamic coefficient $\omega_\alpha$   is  calculated according to Eq.~\eqref{eq:dynPc2}  and plotted in  Fig~\ref{fig4:4dif2}. 
In Fig~\ref{fig4:4dif2}b, we observe that the dynamic coefficient $\omega_\alpha$ is related to the exposure time $\overline{\chi}$ non-monotonically. Thus, it is challenging to propose a  functional relationship between $\omega_\alpha$ and $\overline{\chi}$ directly. 
On the other hand, the dynamic coefficient $\omega_\alpha$ in Fig~\ref{fig4:4dif2}a  is increasing with respect to the exposure time to the WA agent. However, the $\omega_\alpha$-$S$ curves in Fig~\ref{fig4:4dif2}a have piece-wise functional forms i.e., zero and Langmuir-type function of phase saturation, that are altered with exposure time.   This imposes another challenge to find a smooth model to predict the curves along with the saturation. But, one can design a piece-wise function to correlate the $\omega_\alpha-S$ data. From  Fig~\ref{fig4:4dif2}a, we observe that the starting point of the Langmuir functional form is exposure time-dependent along the saturation path. 
Thus, the $\omega_\alpha-S$ can be represented as, 
\begin{equation}\label{omegaalpha}
\omega_\alpha(S_w, \overline{\chi} ):= \left\{ \begin{array}{l}
\frac{S_w-S^*_w(\overline{\chi})}{S_w - S_w^*(\overline{\chi}) + \beta(\overline{\chi})}, ~ {\rm for}~ S_w > S_w^*(\overline{\chi}),\\[0.1in]
0, ~~~~~~~~~~~~~~~{\rm otherwise}
\end{array}\right.
\end{equation}
where $S_w^*$ and $\beta$ are time-dependent. The variable $S_w^*$ is used to transform the starting point of the Langmuir part 
$\omega_\alpha$ along the saturation to zero, and $\beta$ is used to determine the curvature of the curve. From Fig~\ref{fig4:4dif2}a, we observe that the parameters $S_w^*$ and $\beta$ are decreasing functions of exposure time, $\overline{\chi}$. 

We matched the designed model in Eq.~\eqref{omegaalpha} with the dynamic coefficient data, $\omega_\alpha-S_w$, to describe the relations $S_w^*-\overline{\chi}$ and $\beta-\overline{\chi}$. The obtained  relations have the form, 
\begin{equation}\label{paramodel}
S_w^* = a_1\overline{\chi}^{b_1} + c_1,~{\rm and}~\beta = a_2\overline{\chi}^{b_2} + c_3,
\end{equation}
where $a_i$, $b_i$ and $c_i$ for $i=1,2$ are dimensionless fitting parameters. These parameters are estimated and given in Table \ref{TabOmega} for this particular simulation. 
\begin{table}
\centering
\begin{tabular}{l l l l}
\toprule
Parameter & Vale      & Parameter & Value\\
\midrule
$a_1$     & 0.0006123 & $a_2$     & 0.001547\\
$b_1$     & -2.522    & $b_2$     & -2.086\\
$c_1$     & 0.1297    & $c_2$     & 0.2007\\
\bottomrule
\end{tabular}
\caption{The estimated parameter values for the interpolation model.}\label{TabOmega}
\end{table}

\begin{figure}
\centering
\subfigure[]{\includegraphics[scale=0.45]{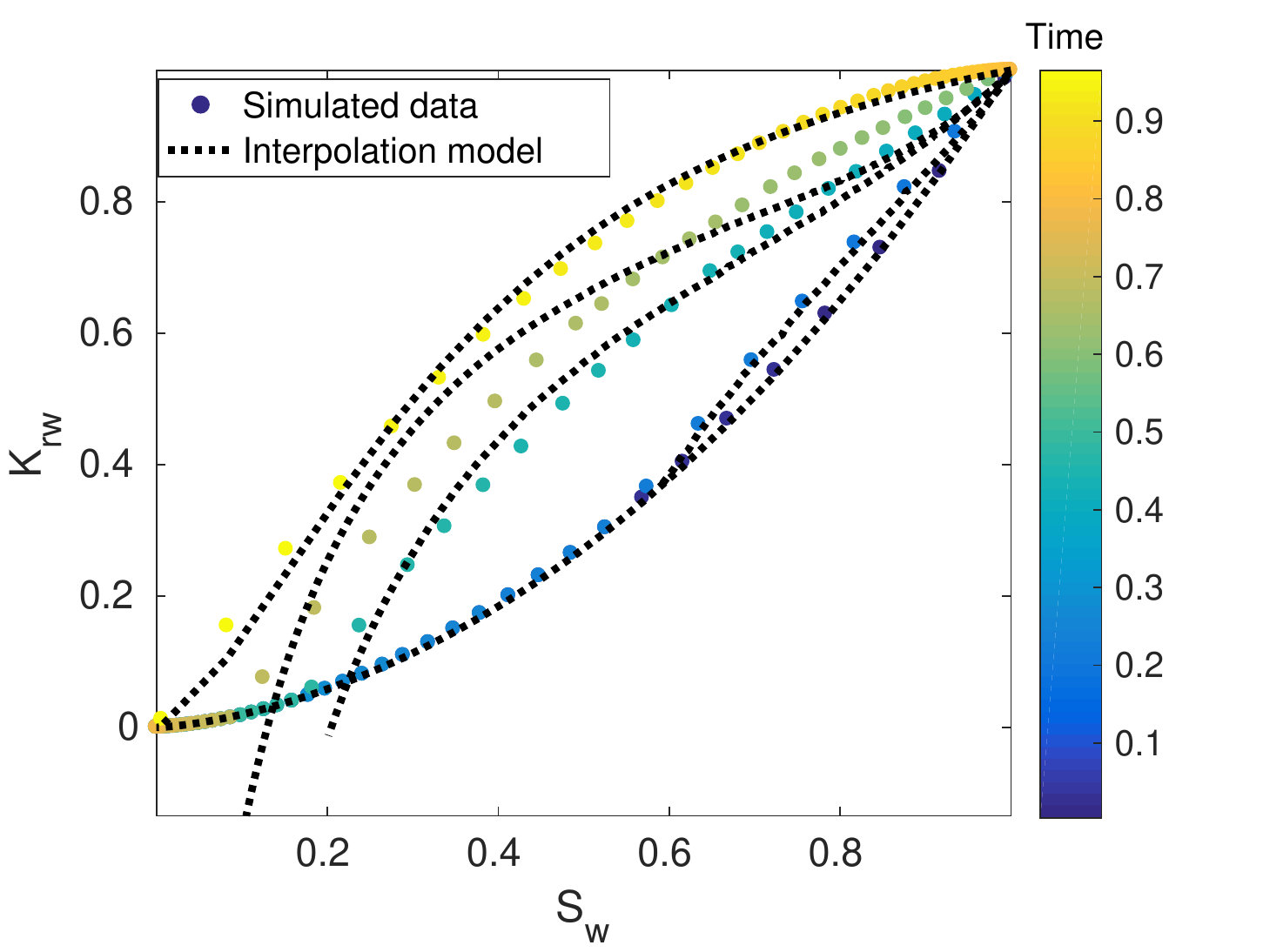}}
\subfigure[]{\includegraphics[scale=0.45]{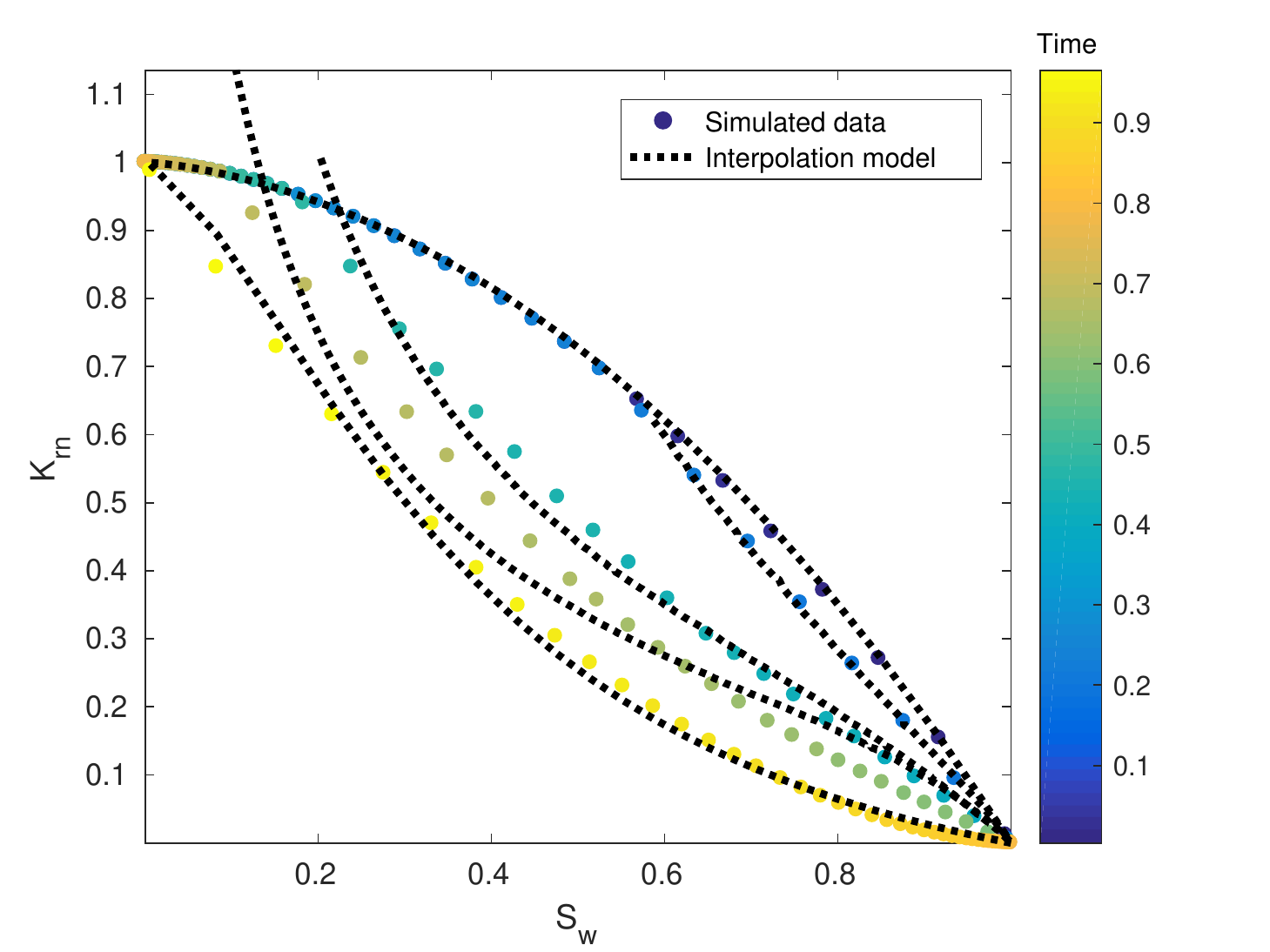}}
\caption{Comparison of the interpolation model \eqref{eq:dynPc_inter} with the simulated (wetting (a) and non-wetting (b)) relative permeabilities. }\label{figInt}
\end{figure}

The models in Eqs. \eqref{omegaalpha}-\eqref{paramodel} are then substituted back into the interpolation model \eqref{eq:dynPc_interp} to give the dynamic relative permeability model and is read as,
\begin{equation}
k_{r\alpha} =\left \{\begin{array}{l}
\frac{S_w-S^*_w(\overline{\chi})}{S_w - S_w^*(\overline{\chi}) + \beta(\overline{\chi})} \big( k_{r\alpha}^{f}-k_{r\alpha}^{i}\big) + k_{r\alpha}^{i},~{\rm for}~ S_w > S^*_w(\overline{\chi})\\[0.1in]
k_{r\alpha}^{i},  ~~~~~~~~~~~~~~~{\rm otherwise}. \label{eq:dynPc_inter}
\end{array}\right.
\end{equation}
The dynamic model \eqref{eq:dynPc_inter} is compared with the simulated relative permeability  in Fig \ref{figInt}. 
According to  Fig \ref{figInt}, the designed model predicts beyond the initial wetting state at joint-point of the initial wetting-state model and the designed interpolation model. This shows that the designed model \eqref{eq:dynPc_inter} is badly correlated with the simulated \krs data. Furthermore,
the non-smoothness behavior of $\omega_\alpha-S_w$  results in a piece-wise phase relative permeability model with many dynamic parameters to be calibrated. 

%
%

\begin{figure}[h!]
\centering
\subfigure[]{\includegraphics[scale=0.45]{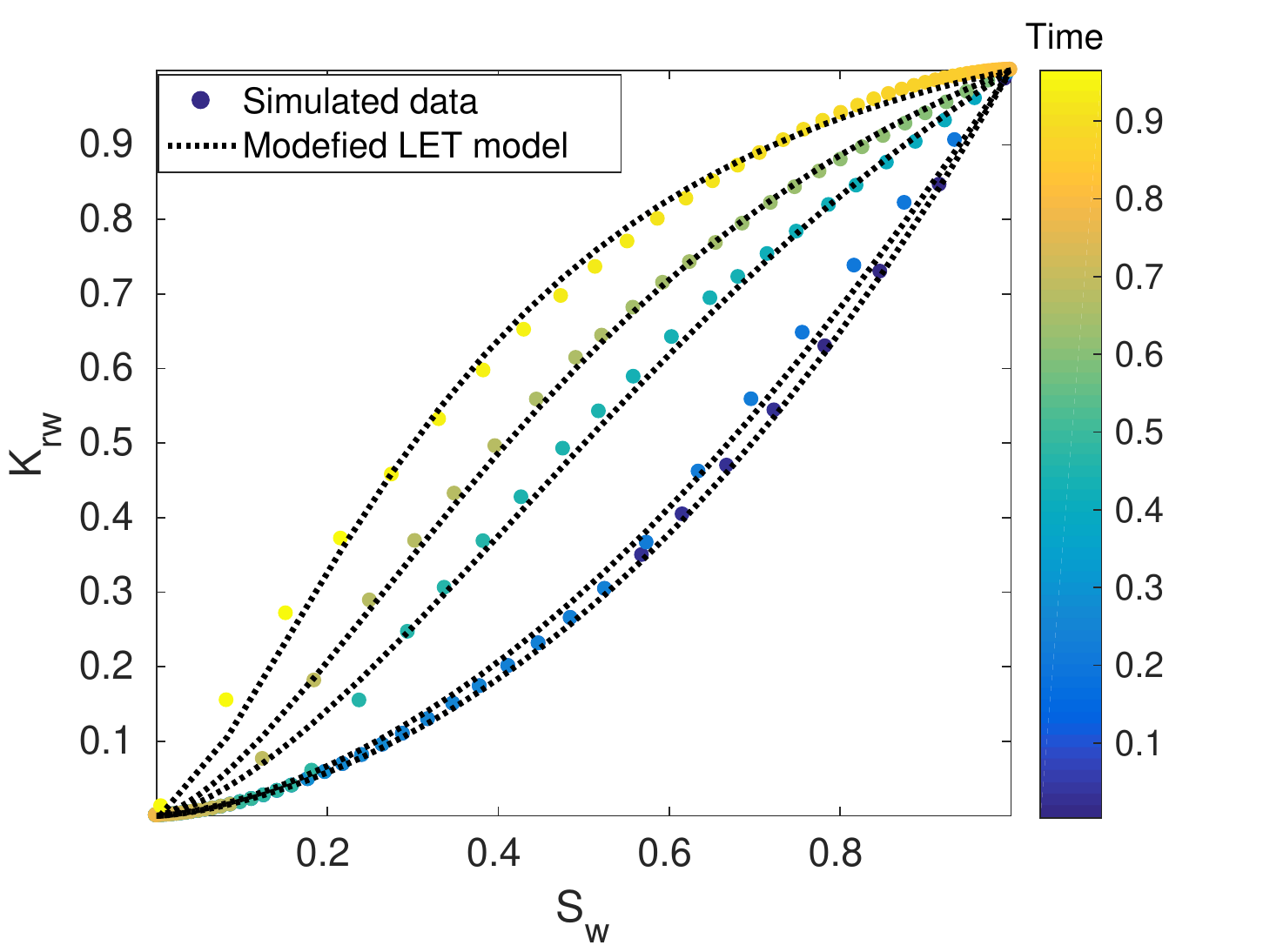}}
\subfigure[]{\includegraphics[scale=0.45]{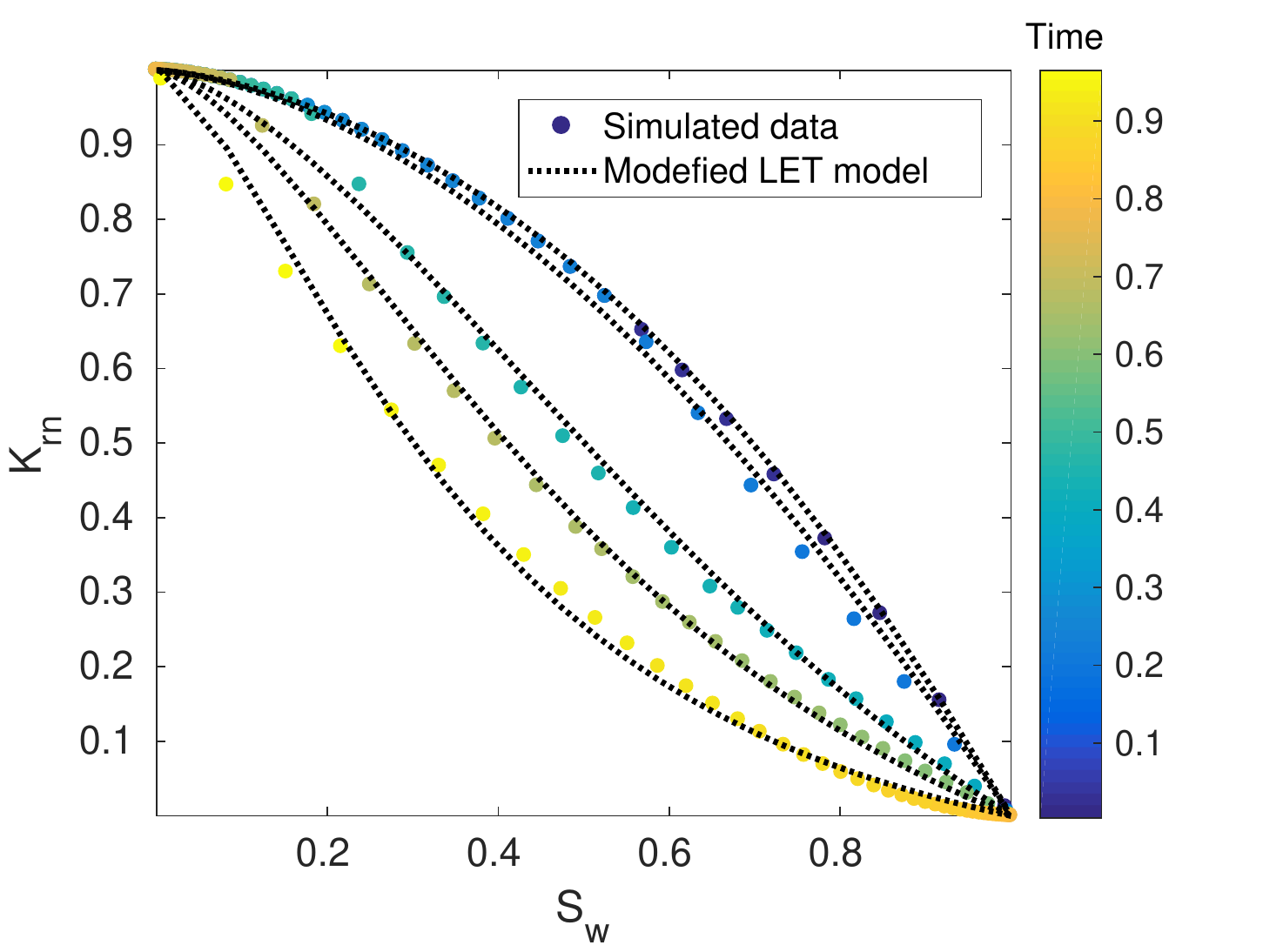}}
\caption{Comparison of the modified LET model and the simulated phase relative permeability: (a) wetting phase and (b) non-wetting phase.}\label{fig:dynaLET}
\end{figure}

The second approach discussed in Section \ref{ModelApp} relies on establishing   a relation between the model parameters and   $\overline{\chi}$ to upscale the effect of pore-scale wettability evolution in relative permeabilities. This is supported  by the result reported in Table \ref{tab:paralist} for end wetting-state curves in which the model  parameters  are dependent on the wetting condition of the porous domain in addition to the pore-size distribution. As a consequence, we can formulate dynamic correlation models from standard  $k_{r\alpha}$-$S$ relations, i.e.,   BC models in Eqs.~\eqref{eq:bc1}-\eqref{eq:bc2} or reduced LET models in Eq.~\eqref{LET_reduced1}. So far, we noticed  that only $E_\alpha$ is sensitive for wettability change in the reduced LET model, whereas both parameters are sensitive in the BC model (see  Table \ref{tab:paralist}). This implies that the BC model is more expensive than the  reduced LET model to upscale the effect of WA on relative permeabilities. Thus, we choose the reduced LET model to represent the dynamic relative permeability curves. Here, we rearrange the reduced LET model as 
\begin{equation}
\label{LET_reduced}
k_{rw} =  \frac{\mathcal{L}_n(\overline{\chi}, E_n) S_w^{\lambda}}{\mathcal{L}_n(\overline{\chi}, E_w) S_w^{\lambda} + 1-S_w},~ {\rm and }~ k_{rn} = \frac{1-S_w}{1-S_w + \mathcal{L}_n(\overline{\chi},E_n) S_w^{\lambda}},
\end{equation}
where $\lambda$ and $E_n$ are determined from the initial wetting-state correlation. Here, $\mathcal{L}_n$ is designed to change with CA. 
Thus, we coupled the reduced LET model \eqref{LET_reduced} with curve fitting tool in MATLAB and matched with the \krs data to study the functional dependencies between the $\mathcal{L}_n$ and $\overline{\chi}$. The obtained  relation is linear and has the form
\begin{equation}\label{Epara}
\mathcal{L}_n(\overline{\chi}, E_n)  = a_n\overline{\chi} + E_n,
\end{equation}
where $E_n$ is as given in  Table \ref{tab:paralist}  and $a_n$ is dynamic fitting parameter for wetting and non-wetting phase relative permeabilities and determines the slope of the relative permeabilities along exposure time. For this particular simulation this parameter is estimated to be $a_n =  3.57$ for all dynamic drainage-imbibition cycles reported above. 
Now the    dynamic term $\mathcal{L}_n$ in Eq.~\eqref{Epara}   can be  substituted  into the reduced LET model \eqref{LET_reduced} to give the dynamic relative permeabilities:
%
%
%
\begin{equation}\label{LETdyna}
k_{rw} =  \frac{(a_n\overline{\chi} + E_n)S_w^{\lambda}}{1-S_w + (a_n\overline{\chi} + E_n) S_w^{\lambda} },~ {\rm and }~ k_{rn} = \frac{1-S_w}{1-S_w + (a_n\overline{\chi} + E_n)S_w^{\lambda}},
\end{equation}
where the parameters $E_n$ and $\lambda$ are pre-determined from the initial wetting-state correlation. 
The proposed dynamic relative permeability models in Eq.~\eqref{LETdyna}  and  the simulated $k_{r\alpha}$-$S$ data are compared in  Fig~\ref{fig:dynaLET}.  
From Fig~\ref{fig:dynaLET}, we observe that the proposed dynamic model \eqref{LETdyna} correlated well with the simulated relative permeability curves. 
The proposed relative permeability model is single-valued regardless of the number of drainage-imbibition cycles. However, this single-valued model is not well predictive around the junction points, particularly for low wetting saturations. However, the obtained correlation result shown in Fig \ref{fig:dynaLET} is acceptable and a significant improvement on the interpolation model result shown in Fig~\ref{figInt}. Furthermore, the modified LET model prediction can be improved by letting the parameter $\lambda$ to vary along the exposure time. However, this may (at least) double the number of parameters in the model that need to be calibrated. 

If we compare the two dynamic models (i.e., the piece-wise interpolation model in Eq.~\eqref{eq:dynPc_inter} and the reduced/modified LET model in Eq.~\eqref{LETdyna}), the reduced LET model is more efficient and simpler to implement in the Darcy flow than the piece-wise interpolation model. Because, the reduced LET model is smooth   along time and saturation.  We also note that the number of parameters in the modified  relative permeability model in Eq.~\eqref{LETdyna} is reduced by half from the original LET model. These all make the reduced dynamic  model more reliable than using a model consisting of multiple parameters that change in each cycle (or hysteresis models). Thus the analysis below will concern only on the modified LET model. 

\subsubsection{The modified LET model sensitivity to pore-scale model parameter} 
\label{sub:sensitivity_to_pore_scale_model_parameter}
\begin{figure}[h!]
\centering
\includegraphics[width=0.48\textwidth]{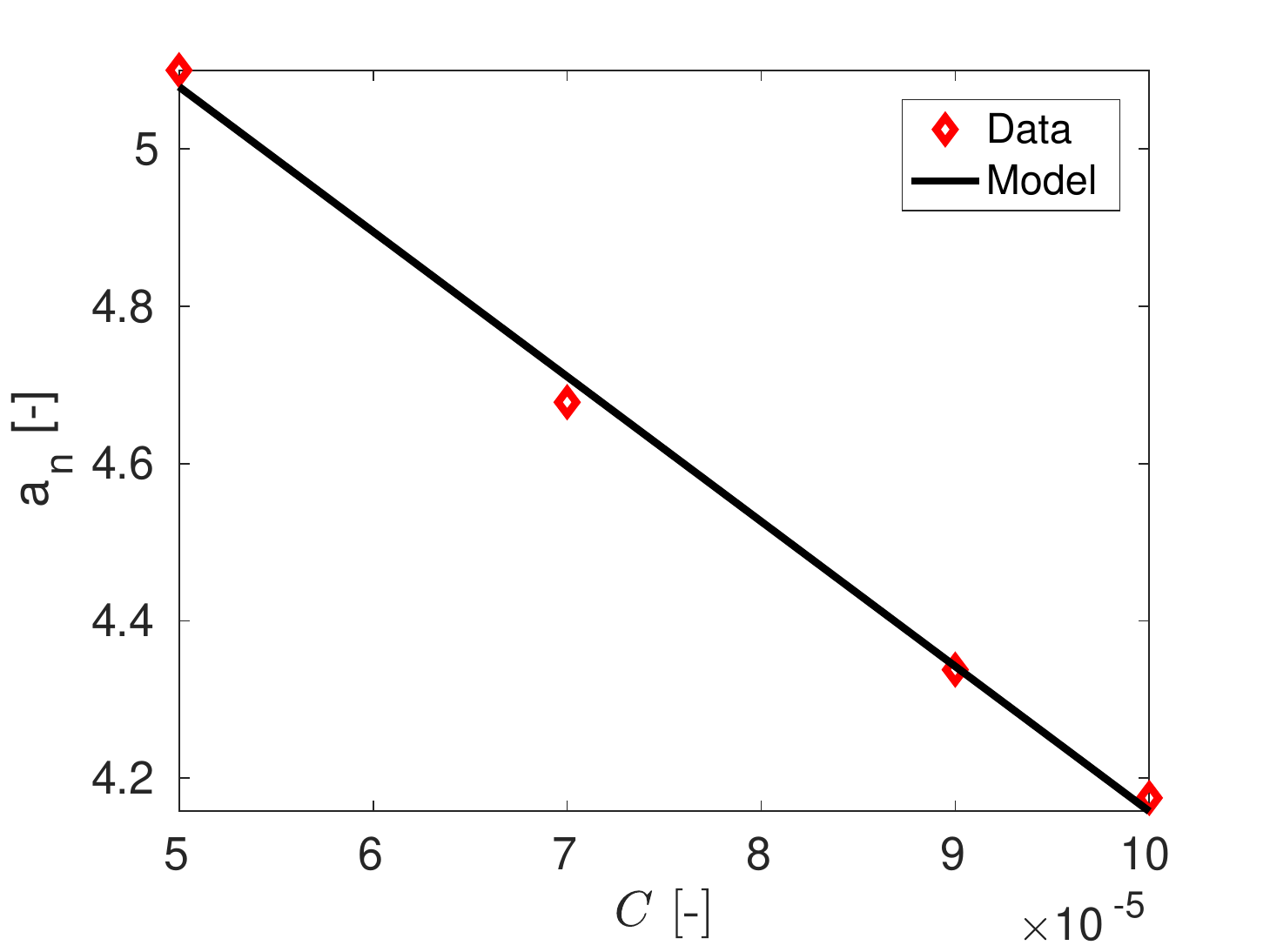} 
\caption{The relation between pore-scale wettability parameter $C$ and the correlation parameter $a_n$ in Eq.~\eqref{LETdyna}.}\label{pararela} 
\end{figure}

In this section, we will investigate the response of  the upscaled model parameter $a_n$ to the change in  the CA model parameter $C$ in Eq.~(\ref{contactGnera}). The parameter $C$ controls the extent  of WA at the pore-level, whereas $a_n$ determines the WA induced dynamics in the relative permeabilities.  We have simulated different drainage-imbibition \krs curves by varying $C$ to draw a relation between $a_n$ and $C$. To do so, we determine  the parameter  $a_n$ from each   \krs data that was simulated by considering   different values of $C$. 
Then, we correlate the estimated parameter values of $a_n$ with the chosen values of $C$, which can be read as
\begin{equation}\label{LinRelBetCandAlpha}
a_n = P_1C+ P_2, 
\end{equation}
where $P_1$ and $P_2$ are fitting parameters. The correlation result is plotted in Fig~\ref{pararela}, where the parameters are estimated to be  $P_1=-18420$ and $P_2 = 6$. Therefore, we can predict the upscaled dynamics of the relative permeabilities directly from the pore-scale WA process.  

The relation in Eq.~\eqref{LinRelBetCandAlpha} can be substituted into the dynamic relative permeability model \eqref{LETdyna} to complete the upscaling process.  
%
The resulting dynamic relative permeability models are saturation, exposure time, and pore-scale WA parameter  dependent. The pore-scale parameter has to be estimated from the calibration of CA model \eqref{contactGnera} with experimental data. Validating the underlying CA change model is beyond the scope of this paper. Rather, we consider the CA model as a reasonable basis to perform and analyze the upscaling process.  


%
\subsection{Applicability of the modified LET model to arbitrary saturation history}
\begin{figure}[h!]
\centering
\subfigure[]{\includegraphics[scale=0.4]{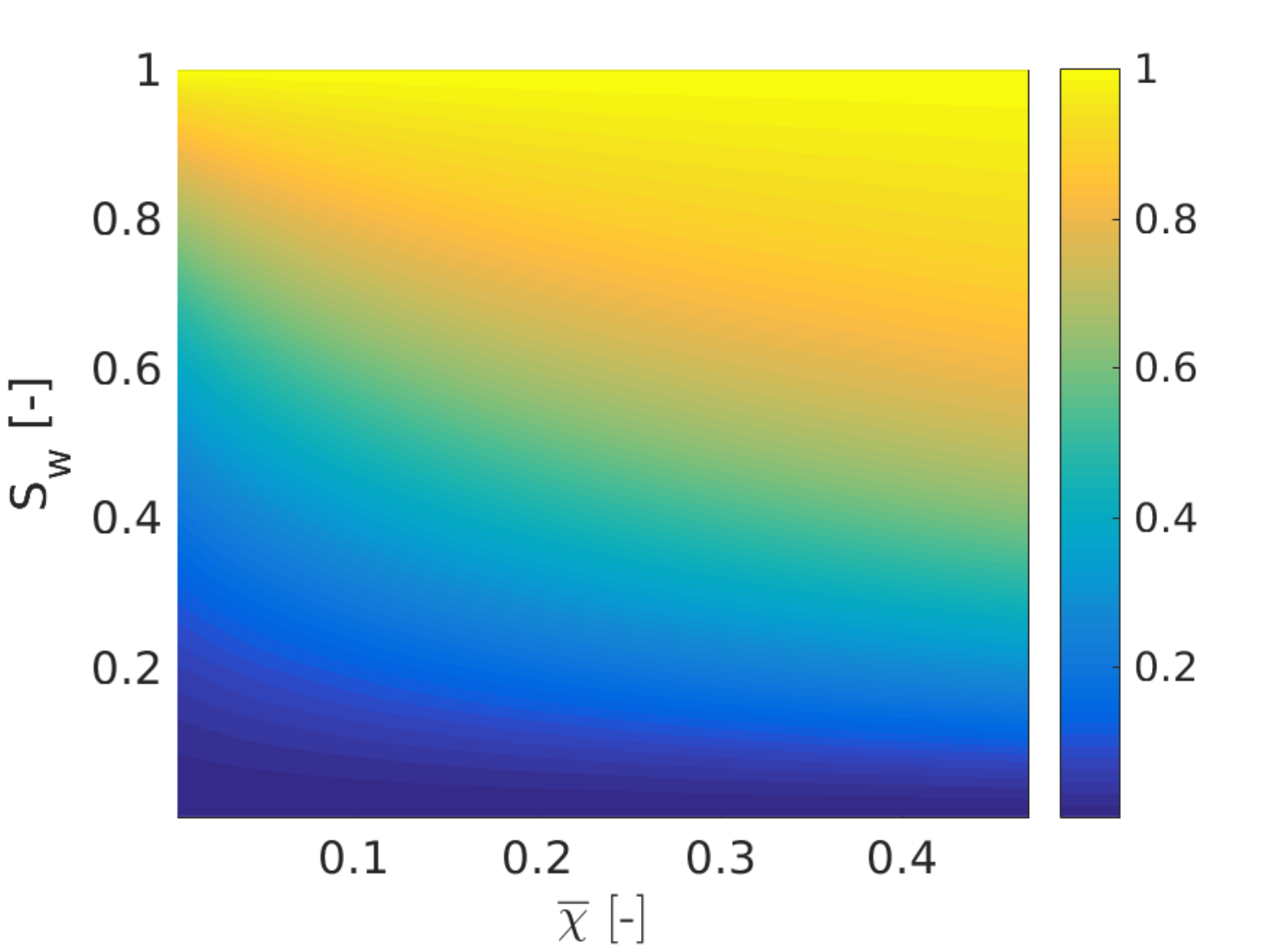}}
\subfigure[]{\includegraphics[scale=0.4]{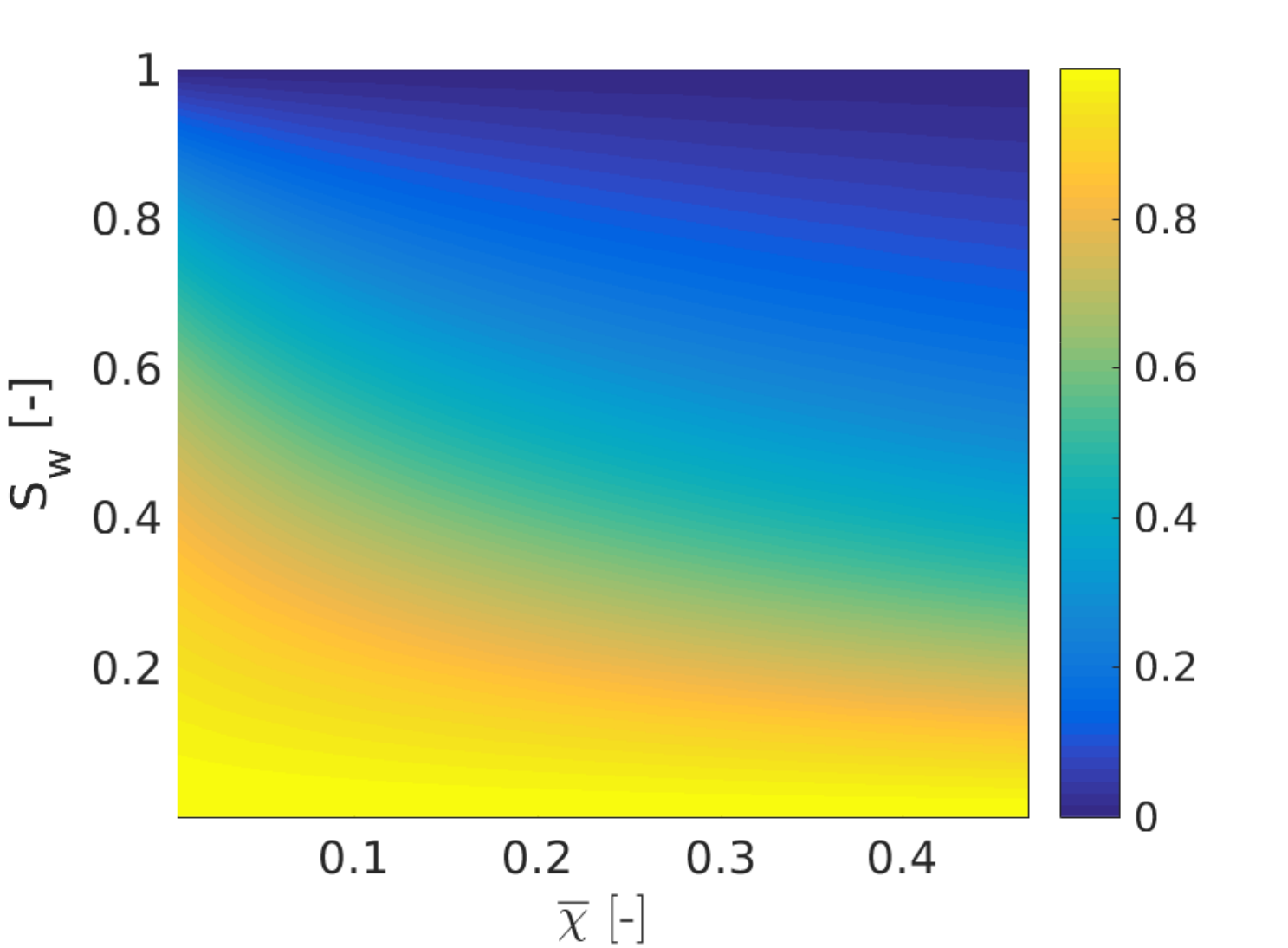}}\\
\subfigure[]{\includegraphics[scale=0.45]{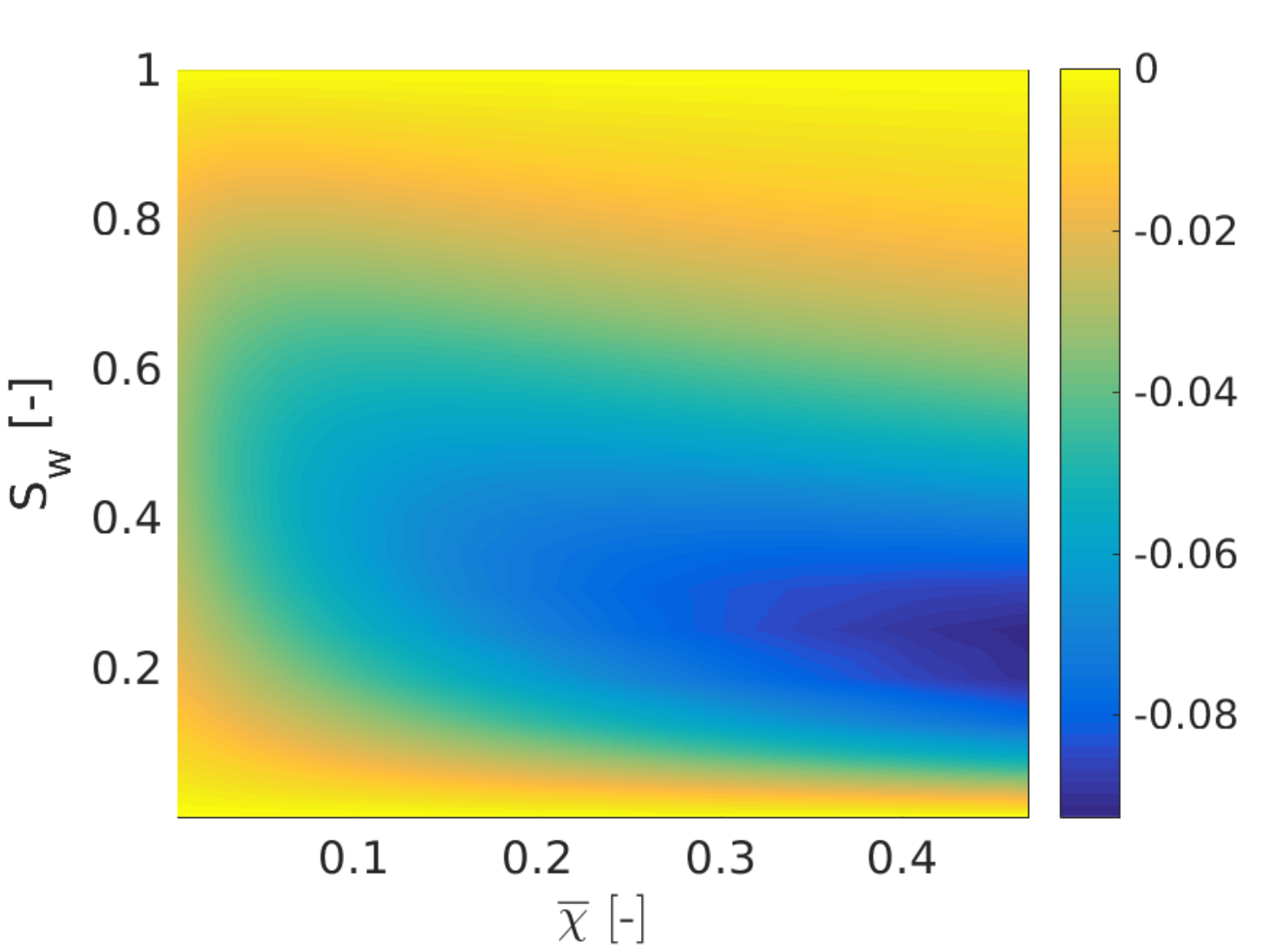}}
\caption{Top: Simulated relative permeabilities obtained by taking multiple paths in the $S_w\times \overline{\chi}$ space, for the wetting  (left) and non-wetting (right) phases. Bottom: The difference between the dynamic model  with the simulated data}\label{surfaceplot}
\end{figure}

The saturation path that used to generate the relative permeability data in Fig~\ref{surfaceplot} can be considered as one of the arbitrary paths in the domain $S_w\times\overline{\chi}$. 
%
This saturation path is dependent on the history of the exposure time to the WA agent (degree of WA change) and the reversal-point of saturation. 
This implies that the relative permeability-saturation dynamics  may behave  differently  if one chooses  a different saturation path that entails a prolonged exposure time for a fixed saturation profile and/or flow reversal at intermediate saturation.  

Here, we simulate as many as possible \krs curves that involve different reversal-points and exposure history within the $S_w\times\overline{\chi}$. Note that we use the pore-scale model parameter $C= 10\times10^{-5}$. The simulated data is plotted in Fig~\ref{surfaceplot}a and \ref{surfaceplot}b for phase relative permeabilities. These arbitrary curves are used to test the potential of the modified LET model in Eq.~\eqref{LETdyna}. To do so, we apply the calibrated dynamic relative permeability model  \eqref{LETdyna} to generate the \krs-$\overline{\chi}$ surface. The absolute difference between the simulated data and the surface generated by the calibrated model is depicted in Fig~\ref{surfaceplot}c. According to the results in Fig~\ref{surfaceplot}, we can justify that a single saturation history path   is sufficient to calibrate a dynamic model that can be applied to any saturation-time path.  
\subsection{Discussion}\label{discussion}
In contrast to the dynamic capillary pressure model presented in \cite{Kassa2019}, the interpolation-based approach is poorly correlated with the simulated relative permeability curves. However, further investigation may be needed to re-evaluate  the potential of capturing the WA process in the relative permeabilities based on the interpolation approach. We also examined the BC model to upscale the WA induced dynamics in the relative permeabilities though we do not report it here fully. For the sake of brevity, we have highlighted only the response of the BC  model  to the wettability change in Section \ref{modeling} for end wetting conditions. In general,   other models that involve more than two parameters (sensitive to CA change) could be calibrated with reasonable accuracy. But, these parameters need to be adjusted in each drainage-imbibition displacements. This complicates the modeling process and the resulting model may involve many parameters that may impose an extra challenge to analyze the WA impact on the flow dynamics at the Darcy scale.

In this study, we have developed a dynamic relative permeability model by modifying an existing static-wettability model i.e., the LET model. The original LET model consists of three parameters in each phase that need to be adjusted differently for different wetting conditions. First, we did numerical experiments to study the dependency of these parameters on the pore-size distribution, while the wettability was kept constant. We found that $T_w$ for wetting and $L_n$ for non-wetting phase relative permeability models are constant for any pore-size distribution. From this, we reduced the LET model to a model (the reduced model in Eq. \eqref{LET_reduced1}) that involves two parameters in each phase. We further investigate the sensitivity of the reduced LET model parameters to the WA dynamics, and we found that only $E_\alpha$ is dependent on CA change dynamics. We then draw a clear relation between $E_\alpha$ and the exposure time $\overline{\chi}$ which allows us to come up with a single-valued dynamic relative permeability model that represents any arbitrary drainage-imbibition cycle. We note that the model involves two types of parameters. The first one is pore-size distribution dependent parameters $E_n$, $L_w$, and $T_n$ which are  determined \textit{a priori} from the initial wetting-state correlation. Knowing these values, the parameter $a_n$ is the only parameter that controls the WA induced dynamics in the relative permeabilities for both phases. 

The proposed model is at the Darcy scale,   that allows for a change in relative permeability as a function of averaged variables such as saturation ($S_w$) and exposure time to a WA agent ($\overline{\chi}$). This implies that the relative permeability in a grid block that is exposed to the WA agent may change over time even for constant  saturation profile. However, if the grid block is not exposed to the WA agent, $\overline{\chi}$ is zero for this particular grid block. In this case, the dynamic model predicts the initial wetting-state curve. However, the developed model may continue further after the end wetting-state curve has attained which is in contrast to the interpolation-type model, where prolonged exposure does not contribute to the dynamics once the final wetting curve is met, see \cite{Kassa2019}.  Thus it is important to propose a strategy to ensure that the relative permeability dynamics do not to cross the end-state curve.  Above, the WA induced dynamics in the relative permeabilities is represented by 
\begin{equation}
\mathcal{L}_n(\overline{\chi}) = a_n \overline{\chi} + E_n
\end{equation}
where $E_n$ is known from the initial wetting-state correlation.  The final wetting state is attained when $a_n \overline{\chi} + E_n = E_n^f$ is satisfied. From this, we can estimate the exposure time needed to reach the final wetting state (say $\overline{\chi}_{\rm max}$) such that the relative permeability is represented by the end wetting-state curve. After knowing this we can set the dynamic variable as
\begin{equation}\label{chicont}
\overline{\chi} = \left\{\begin{array}{l} 
\frac{1}{T}    \int_o^tS_{\rm nw}d\tau, ~{\rm if} ~ \overline{\chi} < \overline{\chi}_{\rm max}\\[0.1in]
\overline{\chi}_{\rm max}, ~{\rm if} ~ \overline{\chi} \geq \overline{\chi}_{\rm max}
\end{array}\right.
\end{equation}
This controls the unnecessary dynamics once the final wetting-state curve is predicted. Nevertheless, the dynamic term in the model pushes the relative permeability towards the higher and lower end of the curve for the wetting and non-wetting phases respectively.

Previous studies \cite{Delshad09, Yu2008, Andersen15, Adibhatla05} represent the impact of instantaneous WA on relative permeabilities by an interpolation model which matched directly to core-scale data in a heuristic manner. This study revealed that the interpolation model is not the best approach to upscale the pore-scale WA process. Rather, we have shown the potential of a modified LET model to capture the underlying WA process at the pore-scale represented by a triangular bundle-of-tubes.  
The proposed models are smooth and simple to use for practical applications. Most importantly, the models are designed to eliminate the relative permeabilities hysteresis induced by CA change during drainage and imbibition displacements.  
 Similar to the developments in \cite{Kassa2019}, we have quantified the link between the pore-scale model parameter $C$ and the core-scale parameter $a_n$. We estimated the core-scale (dynamic) parameter $a_n$ by varying the pore-scale parameter $C$. 
According to the simulation results, we have shown that a very simple scaling can relate a pore-scale process with the core-scale.
This result implies that knowing the mechanism that determines the CA change at the pore-level can be used to predict the macroscale dynamics without performing
pore-scale simulations. This is an important  and valuable generalization for making use of experimental data to inform core-scale relative permeability-saturation relations.


\section{Conclusion} 
In this paper, 
we developed a dynamic relative permeability model that includes the pore-scale underpinnings of WA in the relative permeability--saturation relationships at the Darcy scale. We found that the developed model (i.e., the modified LET model in Eq~\eqref{LET_reduced}) is simple to use and can predict WA induced changes in the relative permeabilities. The modified LET model shows a good agreement with the simulated relative permeability data. Furthermore, this model is independent of the saturation-time paths generated by any drainage-imbibition cycles.   More importantly, the WA dynamics in the relative permeabilities is controlled by a single-valued parameter that has a clear relationship with the  time-dependent CA change model parameter. 

\section*{Acknowledgement}
Funding for this study was through the CHI project (n. 255510) granted through the CLIMIT program of the Research Council of Norway.

\bibliographystyle{plainnat}      
 

\bibliographystyle{numeric}

\end{document}